\documentstyle[12pt]{article}
\newfont{\bbd}{msbm10 scaled\magstep1}
\begin{document}
\thispagestyle{empty}

\def\ve#1{\mid #1\rangle}
\def\vc#1{\langle #1\mid}

\newcommand{\p}[1]{(\ref{#1})}
\newcommand{\be}{\begin{equation}}
\newcommand{\ee}{\end{equation}}
\newcommand{\sect}[1]{\setcounter{equation}{0}\section{#1}}

%%%%%%%%%%%%%%%%
%\renewcommand{\theequation}{\thesection.\arabic{equation}}
%%%%%%%%%%%%%%%%

\newcommand{\vs}[1]{\rule[- #1 mm]{0mm}{#1 mm}}
\newcommand{\hs}[1]{\hspace{#1mm}}
\newcommand{\mb}[1]{\hs{5}\mbox{#1}\hs{5}}
\newcommand{\Db}{{\overline D}}
\newcommand{\bea}{\begin{eqnarray}}

\newcommand{\eea}{\end{eqnarray}}
\newcommand{\wt}[1]{\widetilde{#1}}
\newcommand{\und}[1]{\underline{#1}}
\newcommand{\ov}[1]{\overline{#1}}
\newcommand{\sm}[2]{\frac{\mbox{\footnotesize #1}\vs{-2}}
           {\vs{-2}\mbox{\footnotesize #2}}}
\newcommand{\prt}{\partial}
\newcommand{\eps}{\epsilon}

\newcommand{\R}{\mbox{\rule{0.2mm}{2.8mm}\hspace{-1.5mm} R}}
\newcommand{\Z}{Z\hspace{-2mm}Z}

\newcommand{\cd}{{\cal D}}
\newcommand{\cg}{{\cal G}}
\newcommand{\ck}{{\cal K}}
\newcommand{\cw}{{\cal W}}

\newcommand{\vj}{\vec{J}}
\newcommand{\vl}{\vec{\lambda}}
\newcommand{\vz}{\vec{\sigma}}
\newcommand{\vt}{\vec{\tau}}
\newcommand{\vw}{\vec{W}}
\newcommand{\poiss}{\stackrel{\otimes}{,}}

\def\l#1#2{\raisebox{.2ex}{$\displaystyle
  \mathop{#1}^{{\scriptstyle #2}\rightarrow}$}}
\def\r#1#2{\raisebox{.2ex}{$\displaystyle
 \mathop{#1}^{\leftarrow {\scriptstyle #2}}$}}

%%%%%%%%%%%%%%%%%%%%%%%%%%%%%%%%%%%%%%%%%%%%%%%%%%%

%\begin{document}

\renewcommand{\thefootnote}{\fnsymbol{footnote}}
\newpage
\setcounter{page}{0}
\pagestyle{empty}
\begin{flushright}
{March 2002}\\
{LPENSL--TH--04/2001}
\end{flushright}
\vfill

\begin{center}
{\LARGE {\bf Lax pair formulation of the N=4 Toda chain}}\\[0.3cm]
{\LARGE {\bf (KdV) hierarchy in N=4 superspace}}\\[1cm]

{}~

{\large F. Delduc$^{a,1}$ and A.S. Sorin$^{b,2}$}
{}~\\
\quad \\
{\em {~$~^{(a)}$ Laboratoire de Physique$^\dagger$,
Groupe de Physique Th\'eorique,}}\\
{\em ENS Lyon, 46 All\'ee d'Italie, 69364 Lyon, France}\\[10pt]
{\em {~$~^{(b)}$ Bogoliubov Laboratory of Theoretical Physics, JINR,}}\\
{\em 141980 Dubna, Moscow Region, Russia}~\quad\\

\end{center}

\vfill

{}~

\centerline{{\bf Abstract}}
\noindent
Lax pair and Hamiltonian formulations of the $N=4$ supersymmetric Toda
chain (KdV) hierarchy in $N=4$ superspace are proposed. The general
formulae for the infinite tower of its bosonic flows in terms of the Lax
operator in $N=4$ superspace are derived. A new $N=4$ superfield basis in
which the flows are local is constructed and its embedding in the
$N=4$ $O(4)$ superconformal supercurrent is established. A proof
that the flows possess five complex conjugations and an
infinite-dimensional group of discrete symmetries in $N=4$ superspace is
presented. A relation between the two descriptions of the hierarchy in
$N=4$ superspace used in the literature is established. All known $N=2$
superfield representations of the $N=4$ KdV
hierarchy are shown to derive from our $N=4$ superspace Lax representation.

{}~

{}~



{}~

{}~

\vfill
{\em \noindent
1) E-Mail: francois.delduc@ens-lyon.fr\\
2) E-Mail: sorin@thsun1.jinr.ru }\\
$\dagger$) UMR 5672 du CNRS, associ\'ee \`a l'Ecole Normale Sup\'erieure de
Lyon.
\newpage
\pagestyle{plain}
\renewcommand{\thefootnote}{\arabic{footnote}}
\setcounter{footnote}{0}

\section{Introduction}
Recently, the Lax pair representation of the $N=4$ supersymmetric Toda
chain hierarchy in $N=2$ superspace has been constructed in \cite{dgs}. The
explicit relationship between the $N=4$ supersymmetric Toda chain and KdV
\cite{di, dik} hierarchies has been established in \cite{ds}, and due to
this reason we call both these hierarchies the $N=4$ Toda chain (KdV)
hierarchy. As a byproduct, the relationship, established in \cite{ds},
induces a new Lax pair representation of the $N=4$ KdV hierarchy in $N=2$
superspace. So together with the former Lax pair in $N=2$ superspace
derived in \cite{dg,ik}, two Lax pair representations of the $N=4$ Toda
(KdV) hierarchy in $N=2$ superspace are now known. However, although the
first few flows of the hierarchy are known both in
harmonic \cite{di} and ordinary $N=4$ superspace \cite{dik,ds}, a Lax
pair formulation in $N=4$ superspace is still unknown. Moreover, even a
relation (if any) between the two different descriptions of
the flows in $N=4$ superspace used in \cite{dik} and \cite{ds} is not
established yet.

The present paper addresses both above-stated problems. Thus,
we propose a Lax pair and a Hamiltonian formulation of the $N=4$ Toda
(KdV) hierarchy in $N=4$ superspace and establish a simple relation
between the two above-mentioned descriptions of the hierarchy in $N=4$
superspace. We derive general formulae for its bosonic $N=4$
flows in terms of the Lax operator in $N=4$ superspace and construct an
$N=4$ superfield basis in which they are local. Moreover, we show how to embed
this superfield basis into the $N=4$ $O(4)$ superconformal
supercurrent. We also present a
proof that all these flows possess both an infinite-dimensional
group of discrete Darboux transformations \cite{ls,ols,dgs,ds}
and five complex conjugations in $N=4$ superspace.

\section{Lax pair formulation of the N=4 Toda (KdV) hierarchy
in N=4 superspace}
Our starting point is a manifestly $N=2$ supersymmetric Lax pair
representation of even flows of the $N=4$ Toda chain (KdV) hierarchy
\cite{dgs}
\begin{eqnarray}
L = D_- + v D_{+}^{-1} u,
\label{lax1}
\end{eqnarray}
\begin{eqnarray}
{\textstyle{\partial\over\partial t_l}}L =
[~(L^{2l})_{\ge 0}~,~L~],
\label{laxrepr}
\end{eqnarray}
\begin{eqnarray}
{\textstyle{\partial\over\partial t_l}}v = [(L^{2l})_{\ge 0}v], \quad
{\textstyle{\partial\over\partial t_l}}u =
(-1)^{l+1}[\Big((L^{T})^{2l}\Big)_{\ge 0}u],
\label{flows}
\end{eqnarray}
where the subscript $\ge 0$ denotes the differential part of the operator
and $L^{T}$ is the operator conjugated Lax
operator\footnote{Let us recall
the operator conjugation rules: $D_{\pm}^{T}=-D_{\pm}$,
$(OP)^{T}=(-1)^{d_Od_P}P^{T}O^{T}$, where $O$ ($P$) is an arbitrary
operator with the Grassmann parity $d_O$ ($d_P$), and $d_O$=0 ($d_O=1$)
for bosonic (fermionic) operators $O$. All other rules can be
derived using these. Hereafter, we use the notation $[Of]$ for
an operator $O$ acting on a function $f$.}.
The functions $v\equiv v(z,{\theta}^{+},{\theta}^{-})$ and
$u\equiv u(z,{\theta}^{+},{\theta}^{-})$ are $N=2$ superfields,
and $D_{\pm}$ are fermionic covariant derivatives\footnote{Hereafter, we
explicitly present only nonzero algebra brackets. Throughout this
paper, we shall use the notation $v'\equiv
\partial v \equiv {\partial\over\partial z} v$.}
\begin{eqnarray}
D_{\pm}= \frac{\partial}{\partial {\theta}^{\pm}} +
{\theta}^{\pm} {\partial}, \quad
\{ D_{\pm},D_{\pm}\} = +2{\partial}.
\label{alg}
\end{eqnarray}
It is instructive to rewrite the Lax operator \p{lax1}
in another superfield basis \cite{ds}
\begin{eqnarray}
J\equiv uv + D_-D_+\ln u, \quad
{\overline J}\equiv -uv,
\label{basis}
\end{eqnarray}
where $J\equiv J(z,{\theta}^{+},{\theta}^{-})$ and
${\overline J}\equiv {\overline J}(z,{\theta}^{+},{\theta}^{-})$ are
unconstrained even $N=2$ superfields. It becomes
\begin{eqnarray}
L \equiv D_- - {\overline J} ~\frac{1}{D_{+} +
[D_{-}^{-1} ({\overline J}+J)]}.
\label{lax2}
\end{eqnarray}
We would like to write the flows in $N=4$ superspace. We introduce two
additional odd coordinates $\eta^{\pm}$ and the $N=4$ odd
covariant derivatives
\begin{eqnarray}
&&{\cal D}_{\pm}= \frac{1}{2}(
\frac{\partial}{\partial {\theta}^{\pm}} +
i\frac{\partial}{\partial {\eta}^{\pm}}+
({\theta}^{\pm} +i{\eta}^{\pm}) {\partial}), \quad
{\overline {\cal D}}^{\pm}= \frac{1}{2}(
\frac{\partial}{\partial {\theta}^{\pm}} -
i\frac{\partial}{\partial {\eta}^{\pm}}+
({\theta}^{\pm} -i{\eta}^{\pm}) {\partial}), \nonumber\\
&&\quad \quad \quad \Bigl\{{\cal D}_{k}\,,\,
{\overline {\cal D}}^{m}\Bigr\}=
{{\delta}_{k}}^{m} {\partial}, \quad
\Bigl\{{\cal D}_{k}\,,\,{\cal D}_{m}\Bigr\}=
\Bigl\{{\overline {\cal D}}^{k}\,,\,
{\overline {\cal D}}^{m}\Bigr\}=0,\quad k,m=\pm.
\label{algnn4}
\end{eqnarray}
Our {\it claim} is that
if one replaces the $N=2$ superfields
$J$ and ${\overline J}$ in this basis by one chiral
${\cal J}(z,\theta^+,\theta^-,\eta^+,\eta^-)$ and
one antichiral
${\overline {\cal J}}(z,\theta^+,\theta^-,\eta^+,\eta^-)$
even $N=4$ superfield,
\begin{eqnarray}
{\cal D}_{\pm}{\cal J} =0, \quad
{\overline {\cal D}}^{\pm} ~{\overline {\cal J}} = 0,
\label{N=4constr}
\end{eqnarray}
then the Lax pair representation \p{laxrepr} with the new $N=4$
Lax operator $L_1$,
\begin{eqnarray}
L_1 = D_- - {\overline {\cal J}} ~\frac{1}{D_{+} +
[D_{-}^{-1} ({\overline {\cal J}}+{\cal J})]}, \quad
D_{\pm} \equiv {\cal D}_{\pm}+{\overline {\cal D}}^{\pm},
\label{lax3}
\end{eqnarray}
gives consistent\footnote{We would like to underline that such a
prescription of supersymmetrization leads to inconsistent Lax pair
representations in general except in some cases, one of which is under
consideration.} $N=4$ supersymmetric flows
${\textstyle{\partial\over\partial t_l}}$.
In order to prove
this {\it claim} it is enough to show that the
flow equations \p{flows}, being rewritten in the basis
\p{basis}, are consistent with the
chirality constraints \p{N=4constr}. We present the {\it proof} in a few
steps in what follows.

First, let us simplify the Lax operator $L_1$ \p{lax3} by applying a gauge
transformation and requiring that the gauge transformed Lax
operator
\begin{eqnarray}
{\widetilde L}_1 =  e^{-\xi}L_1e^{\xi}
\label{laxgauge1}
\end{eqnarray}
possesses the following two properties: it should anticommute with the
supersymmetric covariant derivative ${\overline {\cal D}}^{+}$,
\begin{eqnarray}
\{{\overline {\cal D}}^{+}, {\widetilde L}_1\}=0,
\label{laxgauge2}
\end{eqnarray}
and it should contain only a first order term in the inverse fermionic
derivative $D_+^{-1}$. It turns out that these two requirements completely fix
both the gauge transformation function
\begin{eqnarray}
{\xi} \equiv [({\cal D}_- + {\overline {\cal D}}^{-})^{-1}
({\cal D}_+ + {\overline {\cal D}}^{+})^{-1}
({\overline {\cal J}}+{\cal J})]
\label{definition}
\end{eqnarray}
and the form of the Lax operator
\begin{eqnarray}
{\widetilde L}_1 =  {\cal D}_- + {\overline {\cal D}}^{-} +
[({\cal D}_+ + {\overline {\cal D}}^{+})^{-1}
({\overline {\cal J}}+{\cal J})]-
{\overline {\cal J}}{({\cal D}_+ +{\overline {\cal D}}^{+})}^{-1}.
\label{laxrel2g1}
\end{eqnarray}
Then, the Lax pair representation \p{laxrepr} and flows \p{flows} become
\begin{eqnarray}
{\textstyle{\partial\over\partial t_l}}{\widetilde L}_1 =
[~({{\widetilde L}_1}^{2l})_{\ge 0} -
{\textstyle{\partial\over\partial t_l}}\xi~, ~{\widetilde L}_1~],
\label{laxrel2g2}
\end{eqnarray}
\begin{eqnarray}
{(-1)}^l {\textstyle{\partial\over\partial t_l}} {\overline {\cal J}} =
[~{{\Big(\Big(({{\widetilde L}_1}^{T})}^{2l}
\Big)_{\ge 1}\Big)}^{T}~ {\overline {\cal J}}~], \quad
{(-1)}^l {\textstyle{\partial\over\partial t_l}} \xi =
\Big({({{\widetilde L}_1}^{T})}^{2l}\Big)_0,
\label{flowsg}
\end{eqnarray}
respectively, where the subscripts $\ge 1$ and $0$ denote the
differential part of the operator without the nonderivative term and
the nonderivative term, respectively.
After substituting ${\textstyle{\partial\over\partial t_l}} \xi $
from equations \p{flowsg} into \p{laxrel2g2}, the Lax equations become
\begin{eqnarray}
{(-1)}^l {\textstyle{\partial\over\partial t_l}}{\widetilde L}_1 =
\Big[~{{\Big(\Big(({{\widetilde L}_1}^{T})}^{2l}
\Big)_{\ge 1}}\Big)^{T}~, ~{\widetilde L}_1~\Big],
\label{laxrel2g3}
\end{eqnarray}
where we have used the following identity:
\begin{eqnarray}
{(-1)}^l ({{\widetilde L}_1}^{2l})_{+} \equiv
{{\Big(\Big(({{\widetilde L}_1}^{T})}^{2l}
\Big)_{\ge 1}}\Big)^{T} + \Big({({{\widetilde L}_1}^{T})}^{2l}\Big)_{0}.
\label{identity}
\end{eqnarray}
After transposition, \p{laxrel2g3} becomes
\begin{eqnarray}
{(-1)}^{l+1}{\textstyle{\partial\over\partial t_l}}{{\widetilde L}_1}^{T}
= \Big[~{\Big(({{\widetilde L}_1}^{T})}^{2l}
\Big)_{\ge 1}~, ~{{\widetilde L}_1}^{T} ~\Big],
\label{nnn}
\end{eqnarray}
\begin{eqnarray}
{{\widetilde L}_1}^T \equiv
-{\cal D}_- - {\overline {\cal D}}^{-} +
[({\cal D}_+ + {\overline {\cal D}}^{+})^{-1}
({\overline {\cal J}}+{\cal J})]-
{({\cal D}_+ +{\overline {\cal D}}^{+})}^{-1}~{\overline {\cal J}},
\quad  \{{\overline {\cal D}}^{+}, {{\widetilde L}_1}^T\}=0.
\label{laxrel2g1new}
\end{eqnarray}
The transposed Lax operator ${\widetilde L}^{T}_1$ \p{laxrel2g1new}
entering into the Lax pair representation \p{nnn} possesses the
following important properties:
\begin{eqnarray}
(-1)^{l-1}\Big(({\widetilde L}^{T}_1)^{2(l-1)}\Big)_-=
\sum_{k=0}^{2l-3}[({\widetilde L}^{k}_1)^{T}1]
~({\cal D}_+ + {\overline {\cal D}}^{+})^{-1}
~[{\widetilde L}^{2l-3-k}_1{\overline {\cal J}}],
\label{ar}
\end{eqnarray}
\begin{eqnarray}
(-1)^{l}\Big(({\widetilde L}^{T}_1)^{2l-1}\Big)_- & = &
\sum_{k=0}^{2(l-1)}[({\widetilde L}^{k}_1)^{T}1]
~({\cal D}_+ + {\overline {\cal D}}^{+})^{-1}
~[{\widetilde L}^{2(l-1)-k}_1{\overline {\cal J}}] \nonumber\\
&+&\sum_{k=0}^{2l-3}(-1)^{k}[({\widetilde L}^{k}_1)^{T}1]
~({\cal D}_+ + {\overline {\cal D}}^{+})^{-1}
({\cal D}_- + {\overline {\cal D}}^{-})
~[{\widetilde L}^{2l-3-k}_1{\overline {\cal J}}]
\label{ar1}
\end{eqnarray}
which can be derived by rewriting the corresponding formulae of
\cite{dgs} into the superfield basis \p{basis}.

Second, let us discuss some properties of the Lax operator
${{\widetilde L}_1}^T$ \p{laxrel2g1new} which will be useful in what
follows. It can be represented as a sum of two operators
$M$ and ${\overline M}$,
\begin{eqnarray}
&& \quad \quad \quad \quad \quad \quad \quad \quad \quad \quad \quad \quad
{{\widetilde L}_1}^T =M+{\overline M}, \nonumber\\
&& M \equiv -{\cal D}_- +
[~{\overline {\cal D}}^{+}{\partial}^{-1}{\cal J}~], \quad
{\overline M} \equiv  ({\cal D}_+ + {\overline {\cal D}}^{+})^{-1}
\Big({\overline {\cal D}}^{-} -[~{\cal D}_+ {\partial}^{-1}
{\overline {\cal J}}~]\Big)({\cal D}_+ +{\overline {\cal D}}^{+})
\label{prop1}
\end{eqnarray}
with the properties
\begin{eqnarray}
&& M^2 = {\overline M}^2 =0, \quad
\quad  \{{\overline {\cal D}}^{+}, M\}=
\{{\overline {\cal D}}^{+}, {\overline M}\}=0, \quad
\{{\cal D}_{-}, M\}=0, \quad
\{{\overline {\cal D}}^{-}, {\overline M}\}=0, \nonumber\\
&&\{{\cal D}_+,~({\cal D}_+ + {\overline {\cal D}}^{+})^{-1}
~M~({\cal D}_+ + {\overline {\cal D}}^{+})\}=
\{{\cal D}_+,~({\cal D}_+ + {\overline {\cal D}}^{+})
~{\overline M}~({\cal D}_+ + {\overline {\cal D}}^{+})^{-1}\}=0,
\label{prop2}
\end{eqnarray}
\begin{eqnarray}
&& M \equiv ({\cal D}_+ + {\overline {\cal D}}^{+})^{-1}
({\overline M}^{T})^{\#} ({\cal D}_+ + {\overline {\cal D}}^{+}), \quad
{\overline M} \equiv ({\cal D}_+ + {\overline {\cal D}}^{+})^{-1}
(M^{T})^{\#} ({\cal D}_+ + {\overline {\cal D}}^{+}),
\label{prop5}
\end{eqnarray}
where the superscript $\#$ denotes the following substitution (its origin will
be clarified at the end of section 4):
\begin{eqnarray}
\{ {\cal D}_{\pm},{\overline {\cal D}}^{\pm},
{\cal J},{\overline {\cal J}}\}^{\#} =
\{ {\overline {\cal D}}^{\pm}, {\cal D}_{\pm},
-{\overline {\cal J}},-{\cal J}\}
\label{prop6}
\end{eqnarray}
which respects the chirality constraints \p{N=4constr} and which
square is the identity.
Then, one can easily derive some straightforward consequences
\begin{eqnarray}
({{\widetilde L}_1}^T)^2 =\{M,{\overline M}\}
\label{prop3}
\end{eqnarray}
and
\begin{eqnarray}
[M,({{\widetilde L}_1}^T)^{2l} ]= 0, \quad
[{\overline M},({{\widetilde L}_1}^T)^{2l}]=0
\label{prop4}
\end{eqnarray}
as well as
\begin{eqnarray}
({{\widetilde L}_1}^{T})^{\#}=
({\cal D}_+ + {\overline {\cal D}}^{+})^{-1}
{{\widetilde L}_1}({\cal D}_+ + {\overline {\cal D}}^{+}).
\label{prop7}
\end{eqnarray}
We will also need the following relation:
\begin{eqnarray}
\Big(\Big({({{\widetilde L}_1}^{T})}^{2l}\Big)_0\Big)^{\#}=
(-1)^l\Big({({{\widetilde L}_1}^{T})}^{2l}\Big)_0
\label{prop8}
\end{eqnarray}
which results from \p{prop7} and from the identity
\begin{eqnarray}
(D_{+}^{-1}OD_{+})_0=(-1)^{d_O}(O^T)_0
\label{prop9}
\end{eqnarray}
for a pseudo-differential operator $O$ with the Grassmann
parity $d_O$. It is interesting to remark that the square of the
Lax operator ${\widetilde L}^{T}_1$ admits beside \p{prop3} another
representation in the form of an anticommutator,
\begin{eqnarray}
({{\widetilde L}_1}^T)^2 =\{{\overline {\cal D}}^{+},N\}, \quad
N & \equiv & {\cal D}_{+} +[{\overline {\cal D}}^{-}
{\partial}^{-1}{\cal J}]+
({\cal D}_{+} +{\overline {\cal D}}^{+})^{-1}
[{\cal D}_{-}{\partial}^{-1}{\overline {\cal J}}]
({\cal D}_{+} +{\overline {\cal D}}^{+}) \nonumber\\
&-&({\cal D}_{+} +{\overline {\cal D}}^{+})^{-1}{\cal J}
({\cal D}_{+} +{\overline {\cal D}}^{+})^{-1}
[{\cal D}_{+}{\partial}^{-1}{\overline {\cal J}}],
\label{repanti}
\end{eqnarray}
but the square of the operator $N$ does not equal zero as $M$
does, see equation \p{prop2}.
Perhaps one can modify the operator
$N$ to satisfy this property, but we did not succeed in that.

Third, on the basis of the properties
of the Lax operator ${\widetilde L}^{T}_1$ discussed in the previous paragraph,
we are ready to prove the
following three identities:
\begin{eqnarray}
[~{{\Big(\Big(({{\widetilde L}_1}^{T})}^{2l} \Big)_{\ge 1}\Big)}^{T}~
{\overline {\cal J}}~] =
[{\overline {\cal D}}^{+}{\overline {\cal D}}^{-} \Big({({{\widetilde
L}_1}^{T})}^{2l}\Big)_0], \label{relat1}
\end{eqnarray}
\begin{eqnarray}
[{\cal D}_{-} {\overline {\cal D}}^{+}
\Big({({{\widetilde L}_1}^{T})}^{2l}\Big)_0]=0,
\label{relat2}
\end{eqnarray}
\begin{eqnarray}
[{\cal D}_{+} {\overline {\cal D}}^{-}
\Big({({{\widetilde L}_1}^{T})}^{2l}\Big)_0]=0
\label{relat3}
\end{eqnarray}
which are crucial for the {\it proof} of our {\it claim}. Indeed, if they
are satisfied, then it is a simple
exercise to rewrite identically the flows \p{flowsg} in the form
\begin{eqnarray}
(-1)^l {\textstyle{\partial\over\partial t_l}}{\overline {\cal J}}
= [{\overline {\cal D}}^{+}{\overline {\cal D}}^{-}
\Big({({{\widetilde L}_1}^{T})}^{2l}\Big)_0], \quad
(-1)^l {\textstyle{\partial\over\partial t_l}} {\cal J} =
[{\cal D}_{+} {\cal D}_{-} \Big({({{\widetilde L}_1}^{T})}^{2l}\Big)_0]
\label{flowsform}
\end{eqnarray}
which respects manifestly the chirality constraints \p{N=4constr}.
In order to prove the relations (\ref{relat1}--\ref{relat3}), let us
extract the equations resulting from the order $0$ in $D_+$
of the identity
\begin{eqnarray}
\{{\overline {\cal D}}^{+}, ({\widetilde L}^{T}_1)^{2l}\}=0
\label{laxgauge22}
\end{eqnarray}
and from the order $-1$ in $D_+$ of the identities \p{prop4}. They are
\begin{eqnarray}
\Big({({{\widetilde L}_1}^{T})}^{2l}\Big)_{-1}=
[{\overline {\cal D}}^{+}\Big({({{\widetilde L}_1}^{T})}^{2l}\Big)_0],
\label{eq1}
\end{eqnarray}
\begin{eqnarray}
[{\cal D}_{-}\Big({({{\widetilde L}_1}^{T})}^{2l}\Big)_{-1}]=0,
\label{eq2}
\end{eqnarray}
\begin{eqnarray}
-[{\overline {\cal D}}^{-}\Big({({{\widetilde L}_1}^{T})}^{2l}\Big)_{-1}]=
[~{{\Big(\Big(({{\widetilde L}_1}^{T})}^{2l} \Big)_{\ge 1}\Big)}^{T}~
{\overline {\cal J}}~],
\label{eq3}
\end{eqnarray}
respectively, where the subscript $-1$ denotes the coefficient of the
inverse derivative $D^{-1}_+$ of a pseudodifferential operator. Now, the quantity
$\Big({({{\widetilde L}_1}^{T})}^{2l}\Big)_{-1}$ being substituted
from equation \p{eq1} into equations (\ref{eq2}--\ref{eq3})
just leads to the relations (\ref{relat1}--\ref{relat2}). As concerns
the remaining relation \p{relat3}, it can easily be derived from
the relation \p{relat2} if the substitution \p{prop6} is applied
to it and then the identity \p{prop8} is used. Alternatively,
it can be obtained starting from the identities
\begin{eqnarray}
\{{\cal D}_{+},
~\Big(({\cal D}_{+}+{\overline {\cal D}}^{+})^{-1}
~{\widetilde L}_1
~({\cal D}_{+}+{\overline {\cal D}}^{+})\Big)^{2l}\}=0
\label{another1}
\end{eqnarray}
and
\begin{eqnarray}
\{({\cal D}_{+}+{\overline {\cal D}}^{+})^{-1}
~{\overline M}^{T}~({\cal D}_{+}+{\overline {\cal D}}^{+}),
~\Big(({\cal D}^{+}+{\overline {\cal D}}^{+})^{-1}
~{\widetilde L}_1
~({\cal D}_{+}+{\overline {\cal D}}^{+})\Big)^{2l}\}=0,
\label{another2}
\end{eqnarray}
which are related in an obvious way to the identities \p{laxgauge22} and
\p{prop4}. Restricting to the orders $0$ and $-1$, respectively, one obtains
\begin{eqnarray}
~\Big(\Big(({\cal D}_{+}+{\overline {\cal D}}^{+})^{-1}
~{\widetilde L}_1
~({\cal D}_{+}+{\overline {\cal D}}^{+})\Big)^{2l}\Big)_{-1}=
[{\cal D}_{+}~\Big(\Big(({\cal D}_{+}+{\overline {\cal D}}^{+})^{-1}
~{\widetilde L}_1
~({\cal D}_{+}+{\overline {\cal D}}^{+})\Big)^{2l}\Big)_0],
\label{another3}
\end{eqnarray}
and
\begin{eqnarray}
[{\overline {\cal D}}^{-}
~\Big(\Big(({\cal D}_{+}+{\overline {\cal D}}^{+})^{-1}
~{\widetilde L}_1
~({\cal D}_{+}+{\overline {\cal D}}^{+})\Big)^{2l}\Big)_{-1}]=0.
\label{another4}
\end{eqnarray}
Then we substitute the quantity $\Big(\Big(({\cal D}_{+}+
{\overline {\cal D}}^{+})^{-1} ~{\widetilde L}_1
~({\cal D}_{+}+{\overline {\cal D}}^{+})\Big)^{2l}\Big)_{-1}$ from
\p{another3} into \p{another4},
\begin{eqnarray}
[{\cal D}_+ {\overline {\cal D}}^{-}
~\Big(({\cal D}_{+}+{\overline {\cal D}}^{+})^{-1}
~{{\widetilde L}_1}^{2l}
~({\cal D}_{+}+{\overline {\cal D}}^{+})\Big)_{0}]=0,
\label{another5}
\end{eqnarray}
and, at last, use the identity \p{prop9}. This ends the {\it proof}
of our {\it claim}.

One important remark is in order. The general resolution of
the identities (\ref{relat2}--\ref{relat3}) and \p{prop8} leads to the following
general representation for the quantity
$\Big({({{\widetilde L}_1}^{T})}^{2l}\Big)_0$:
\begin{eqnarray}
\Big({({{\widetilde L}_1}^{T})}^{2l}\Big)_0 =
[{\cal D}_{+}{\cal D}_{-}X_l]+(-1)^l
~[{\overline {\cal D}}^{+}{\overline {\cal D}}^{-}X^{\#}_l]+
[({\cal D}_{-}{\overline {\cal D}}^{-}-
{\overline {\cal D}}^{+}{\cal D}_{+})~Y_l], \quad
Y^{\#}_l=(-1)^{l+1}Y_l,
\label{relatgeneral}
\end{eqnarray}
where $X_l$ and $Y_l$ are $N=4$ unconstrained nonlocal
functionals of the basic superfields.
Actually, the functionals $X_l$ and $Y_l$ are
ambiguously defined because both of them are acted upon by the
projectors ${\cal D}_{+}{\cal D}_{-}$ and
${\overline {\cal D}}^{+}{\overline {\cal D}}^{-}$,
which can be expressed
via the operator ${\cal D}_{-}{\overline {\cal D}}^{-}-
{\overline {\cal D}}^{+}{\cal D}_{+}$ as
\begin{eqnarray}
{\cal D}_{+}{\cal D}_{-}\equiv
({\cal D}_{-}{\overline {\cal D}}^{-}- {\overline {\cal D}}^{+}
{\cal D}_{+}){\partial}^{-1}{\cal D}_{+}{\cal D}_{-}, \quad
{\overline {\cal D}}^{+}{\overline {\cal D}}^{-}\equiv
-({\cal D}_{-}{\overline {\cal D}}^{-}- {\overline {\cal D}}^{+}
{\cal D}_{+}){\partial}^{-1}{\overline {\cal D}}^{+}
{\overline {\cal D}}^{-}.
\label{relatgeneral2}
\end{eqnarray}
Let us present, as an example, explicit expressions for $X_l$ and $Y_l$ for $l=1$ and
$l=2$,
\begin{eqnarray}
&&X_1=[{\partial}^{-1}{\overline {\cal J}}], \quad
X_2=-{\overline {\cal J}}+
[{\cal D}_{-}{\partial}^{-1}{\overline {\cal J}}]
[{\cal D}_{+}{\partial}^{-1}{\overline {\cal J}}], \nonumber\\
&& Y_1=0, \quad \quad \quad \quad
Y_2=[{\cal D}_{-}{\partial}^{-1}{\overline {\cal J}}]
[{\overline {\cal D}}^{-}{\partial}^{-1}{\cal J}]+
[{\cal D}_{+}{\partial}^{-1}{\overline {\cal J}}]
[{\overline {\cal D}}^{+}{\partial}^{-1}{\cal J}].
\label{relatgeneral1}
\end{eqnarray}

As a next remark, let us show that one can give two different presentations of the
flows \p{nnn} in $N=4$ superspace which are reminiscent of the formalisms used in
\cite{dg} and in \cite{ks3,dg3} in the N=2 supersymmetric case.
First, multiplying the Lax pair representation
\p{nnn} on both sides by the projector
${\overline {\cal D}}^{+}{\cal D}_+{\partial}^{-1}$,
it  becomes
\begin{eqnarray}
{(-1)}^{l+1}{\textstyle{\partial\over\partial t_l}}{\hbox{\bbd L}} =
[~({{\hbox{\bbd L}}}^{2l})_{+}~,~{\hbox{\bbd L}}~],
\label{laxreprnew2}
\end{eqnarray}
where ${\hbox{\bbd L}}$ is the chirality preserving Lax operator,
\begin{eqnarray}
{\hbox{\bbd L}} & \equiv & {\overline {\cal D}}^{+}{\cal D}_+
{\partial}^{-1}
~{\widetilde L}^{T}_1~{\overline {\cal D}}^{+}{\cal D}_+{\partial}^{-1}
\equiv {\overline {\cal D}}^{+}{\cal D}_+{\partial}^{-1}
(-{\cal D}_- - {\overline {\cal D}}^{-}+
[({\cal D}_+ + {\overline {\cal D}}^{+})^{-1}
({\overline {\cal J}}+{\cal J})])
{\overline {\cal D}}^{+}{\cal D}_+{\partial}^{-1} \nonumber\\
& \equiv &  -{\overline {\cal D}}^{+}{\cal D}_+{\partial}^{-1}
~e^{\xi}~({\cal D}_- + {\overline {\cal D}}^{-})~e^{-\xi}~
{\overline {\cal D}}^{+}{\cal D}_+{\partial}^{-1}, \quad
{\overline {\cal D}}^{+}{\hbox{\bbd L}}=
{\hbox{\bbd L}}{\cal D}_+ =0,
\label{newlax}
\end{eqnarray}
and the subscript $+$ denotes the part of the operator
${{\hbox{\bbd L}}}^{2l}$ which can be represented in the form
${\overline {\cal D}^{+}}{\cal O}{\cal D}_+$, where
${\cal O}$ is a differential operator \cite{dg}. In order to show
that the flows \p{laxreprnew2} coincide with those in \p{nnn},
one has to make use of two properties of the operator
${\widetilde L}^{T}_1$. First, its even powers are
pseudo-differential operators containing powers of the derivative
$D_+=\overline {\cal D}^{+}+{\cal D}_+$ only. Second, the operator
${\widetilde L}^{T}_1$ commutes with $\overline {\cal D}^{+}$.
This presentation will be useful when we will study an alternative
Lax pair construction in section 6.

As a second, distinct presentation of the flows \p{nnn}, let us introduce
the operator
\begin{equation}{\cal L}=D_++[D_-^{-1}(\overline{\cal J}+{\cal J})]
=e^{\xi}D_+e^{-\xi}.\end{equation}
It may be used to rewrite the operator $L_1$ in \p{lax3} as
\begin{equation}
L_1={\cal D}_--{\cal L}\overline{\cal D}^-{\cal L}^{-1}.
\end{equation}
Then, the square of the Lax operator $L_1$ reads
\begin{equation}
L_1^2=-\{ {\cal D}_-,{\cal L}\overline{\cal D}_-{\cal L}^{-1}\}
\Rightarrow [{\cal D}_-, L_1^2]=0.\end{equation}
We may define an alternative Lax operator by
\begin{equation}
{\hat L}_1={\cal L}^{-1}L_1{\cal L}.
\label{twolax}\end{equation}
Its square reads
\begin{equation}
{\hat L}_1^2=-\{ \overline{\cal D}_-,{\cal L}^{-1}{\cal D}_-{\cal L}\}
\Rightarrow [\overline{\cal D}_-, L_1^2]=0.\end{equation}
As a consequence of \p{twolax}, one has the equation
\begin{equation}
L_1^{2l}{\cal L}-{\cal L}{\hat L}_1^{2l}=0.
\end{equation}
Then, in the spirit of references \cite{ks3,dg3}, we consider the flow
equations
\begin{equation}{\partial\over\partial t_l}{\cal L}=
(L_1^{2l})_{\geq 0}{\cal L}-{\cal L}({\hat L}_1^{2l})_{\geq 0},
\label{prelax}\end{equation}
which do not have the Lax form. However, it is a matter of simple calculations to
show that, as a consequence of \p{prelax}, the operator $L_1$ satisfies
the Lax equations \p{nnn}. Let us note that, by considering a differential
operator ${\cal L}$ of higher order, one may hope to generalize the N=4 hierarchy
considered in this article.

The $N=4$ Toda (KdV) hierarchy flows
\p{flowsform} in the $N=4$ superfield basis
$\{{\cal J},{\overline {\cal J}}\}$ are nonlocal because of the
nonlocal dependence of the Lax operator ${\widetilde L}^{T}_1$
\p{laxrel2g1new} on the superfields ${\cal J}$ and
${\overline {\cal J}}$. Nevertheless, it is possible to obtain
local flows by introducing a new superfield basis
$\{{\Omega}, {\overline {\Omega}}\}$ defined by
the following invertible transformations:
\begin{eqnarray}
&&{\cal J} \equiv {\cal D}_{+}{\overline \Omega}, \quad \quad
~{\overline {\cal J}}\equiv {\overline {\cal D}}^{+}\Omega,\nonumber\\
&&{\overline \Omega}\equiv {\overline {\cal D}}^{+}
{\partial}^{-1}{\cal J},
\quad \Omega \equiv {\cal D}_{+} {\partial}^{-1}{\overline {\cal J}},
\label{basisnew1}
\end{eqnarray}
where $ \Omega $ and ${\overline \Omega}$ are new constrained odd $N=4$
superfields
\begin{eqnarray}
{\cal D}_{+}{\Omega} ={\overline {\cal D}}^{-}\Omega =0, \quad
{\cal D}_{-} {\overline {\Omega}} =
{\overline {\cal D}}^{+} {\overline {\Omega}} = 0.
\label{basisnew2}
\end{eqnarray}
Then, in terms of the superfields $\{{\Omega}, {\overline {\Omega}}\}$
the Lax operator ${\widetilde L}^{T}_1$ \p{laxrel2g1new} becomes local,
\begin{eqnarray}
{{\widetilde L}_1}^T \equiv
-{\cal D}_- - {\overline {\cal D}}^{-} +
{\Omega}+{\overline {\Omega}}-
{({\cal D}_+ +{\overline {\cal D}}^{+})}^{-1}~
[{\overline {\cal D}}^{+}{\Omega}].
\label{basisnew3}
\end{eqnarray}
Using the identities (\ref{relat2}--\ref{relat3}) one can easily
rewrite the flows \p{flowsform} in this basis as well,
\begin{eqnarray}
(-1)^l {\textstyle{\partial\over\partial t_l}}{\Omega}
= [{\overline {\cal D}}^{-}
\Big({({{\widetilde L}_1}^{T})}^{2l}\Big)_0], \quad
(-1)^l {\textstyle{\partial\over\partial t_l}} {\overline {\Omega}} =
[{\cal D}_{-} \Big({({{\widetilde L}_1}^{T})}^{2l}\Big)_0],
\label{basisnew4}
\end{eqnarray}
where they are obviously local. Actually, beside the basis \p{basisnew1}
there are also three other superfield bases with constrained odd
$N=4$ superfields $ \{{\Psi}, {\overline \Psi}\}$,
$ \{{\Sigma}, {\overline \Sigma}\}$ and $ \{{\Xi}, {\overline \Xi}\}$,
which provide local flows, although the reason for this is
less evident than for the basis \p{basisnew1}. The corresponding
formulae are
\begin{eqnarray}
&&{\overline \Omega}= {\overline \Psi}, \quad {\Omega}\equiv
{\cal D}_{+}{\overline {\cal D}}^{-}{\partial}^{-1}\Psi, \quad
{\cal D}_{-}{\Psi} ={\overline {\cal D}}^{+}\Psi =0, \quad
{\cal D}_{-} {\overline {\Psi}} =
{\overline {\cal D}}^{+} {\overline {\Psi}} = 0, \nonumber\\
&&{{\widetilde L}_1}^T =
e^{\psi} \Big(-{\cal D}_- - {\overline {\cal D}}^{-} +
{\overline {\Psi}}-
\frac{1}{{\cal D}_+ +{\overline {\cal D}}^{+}-\Psi}
~[{\overline {\cal D}}^{-}{\Psi}]~\Big)e^{-\psi}, \quad
\psi \equiv -[{\cal D}_{+}{\partial}^{-1}\Psi], \nonumber\\
&& \quad \quad \quad \quad
(-1)^{l+1} {\textstyle{\partial\over\partial t_l}}{\Psi}
= [{\overline {\cal D}}^{+}
\Big({({{\widetilde L}_1}^{T})}^{2l}\Big)_0], \quad
(-1)^l {\textstyle{\partial\over\partial t_l}} {\overline {\Psi}} =
[{\cal D}_{-} \Big({({{\widetilde L}_1}^{T})}^{2l}\Big)_0].
\label{basisnew6}
\end{eqnarray}
\begin{eqnarray}
&&{\overline \Omega}=
{\cal D}_{-}{\overline {\cal D}}^{+}{\partial}^{-1}{\overline \Sigma},
\quad {\Omega}\equiv \Sigma, \quad
{\cal D}_{+}{\Sigma} ={\overline {\cal D}}^{-}\Sigma =0, \quad
{\cal D}_{+} {\overline {\Sigma}} =
{\overline {\cal D}}^{-} {\overline {\Sigma}} = 0, \nonumber\\
&&{{\widetilde L}_1}^T =
e^{\sigma} \Big(-{\cal D}_- - {\overline {\cal D}}^{-} +{\Xi}-
\frac{1}{{\cal D}_+ +{\overline {\cal D}}^{+}+{\overline {\Sigma}}}
~[{\overline {\cal D}}^{+}{\Sigma}]~\Big)e^{-\sigma}, \quad
\sigma \equiv [{\overline {\cal D}}^{+}{\partial}^{-1}
{\overline \Sigma}], \nonumber\\ && \quad \quad \quad \quad
(-1)^{l} {\textstyle{\partial\over\partial t_l}}{\Sigma}
= [{\overline {\cal D}}^{-}
\Big({({{\widetilde L}_1}^{T})}^{2l}\Big)_0], \quad
(-1)^l {\textstyle{\partial\over\partial t_l}} {\overline {\Sigma}} =
[{\cal D}_{+} \Big({({{\widetilde L}_1}^{T})}^{2l}\Big)_0].
\label{b3}
\end{eqnarray}
\begin{eqnarray}
&&{\overline \Omega}=
{\cal D}_{-}{\overline {\cal D}}^{+}{\partial}^{-1}{\overline \Xi}, \quad
{\Omega}\equiv {\cal D}_{+}{\overline {\cal D}}^{-}
{\partial}^{-1}\Xi,\quad
{\cal D}_{-}{\Xi} ={\overline {\cal D}}^{+}\Xi =0, \quad
{\cal D}_{+} {\overline {\Xi}} =
{\overline {\cal D}}^{-} {\overline {\Xi}} = 0, \nonumber\\
&&\quad \quad \quad \quad \quad \quad \quad \quad \quad \quad \quad \quad
\quad \quad \quad
{{\widetilde L}_1}^T =
e^{\xi} L^{T}_1e^{-\xi},
\nonumber\\ && \quad \quad \quad \quad
(-1)^{l+1} {\textstyle{\partial\over\partial t_l}}{\Xi}
= [{\overline {\cal D}}^{+}
\Big({({{\widetilde L}_1}^{T})}^{2l}\Big)_0], \quad
(-1)^l {\textstyle{\partial\over\partial t_l}} {\overline {\Xi}} =
[{\cal D}_{+} \Big({({{\widetilde L}_1}^{T})}^{2l}\Big)_0],
\label{b4}
\end{eqnarray}
where $L^{T}_1$ and $\xi$ in equations \p{b4} are defined by
\p{lax3} and \p{definition}, respectively, and become in this basis
\begin{eqnarray}
L^{T}_1 =-{\cal D}_- - {\overline {\cal D}}^{-} -
\frac{1}{{\cal D}_+ +{\overline {\cal D}}^{+}+{\overline {\Xi}}-{\Xi}}
~[{\overline {\cal D}}^{-}{\Xi}], \quad
\xi \equiv [{\overline {\cal D}}^{+}{\partial}^{-1}{\overline \Xi}]-
[{\cal D}_{+}{\partial}^{-1}\Xi].
\label{b5}
\end{eqnarray}
Although the Lax operator ${{\widetilde L}_1}^{T}$ is nonlocal
in terms of the superfields involved for each of these three bases, the quantity
$\Big(({{{\widetilde L}_1}^{T})}^{2l}\Big)_0$ entering into the
corresponding equations (\ref{basisnew6}--\ref{b4}) is in fact local.
Indeed, let us demonstrate this
fact for the quantity $\Big(({{{\widetilde L}_1}^{T})}^{2l}\Big)_0$,
e.g., for the basis \p{b4}. In this case,
\begin{eqnarray}
\Big({({{\widetilde L}_1}^{T})}^{2l}\Big)_0 \equiv
\Big({(L^{T}_1)}^{2l}\Big)_0 +
\Big(e^{\xi}\Big({(L^{T}_1)}^{2l}\Big)_{\ge 1}e^{-\xi}\Big)_0.
\label{b6}
\end{eqnarray}
The first term in the right-hand side of equation \p{b6} is local
because of locality of the Lax operator $L^{T}_1$ \p{b5} with
respect to the superfields $\Xi$ and ${\overline \Xi}$.
So, nonlocality could only come from the second term, due to
nonlocality of $\xi$ \p{b5}. However, this term is actually a polynomial
in derivatives of $\xi$, which are local due to the chirality
properties \p{b4} of the superfields $\Xi$
and ${\overline \Xi}$. Therefore, all potential nonlocalities in fact
disappear. The same argument is valid with respect to
each basis from the set (\ref{basisnew6}--\ref{b4}).

As an example, we would like to explicitly present a few first nontrivial
manifestly $N=4$ supersymmetric flows resulting from equations \p{flowsform}
and \p{basisnew6},
\begin{eqnarray}
&&{\textstyle{\partial\over\partial t_2}} {\cal J} =
-{\cal J}~'' -{\cal D}_{+}{\cal D}_{-}\Bigl[2({\cal J} {\partial}^{-1}
{\overline {\cal J} })~'- ({\overline {\cal D}}^{+}
{\overline {\cal D}}^{-}{\partial}^{-1}{\cal J} )^2\Bigr], \nonumber\\
&&{\textstyle{\partial\over\partial t_2}}{\overline {\cal J}} =
+{\overline {\cal J}}~''  - {\overline {\cal D}}^{+}
{\overline {\cal D}}^{-}
\Bigl[2({\overline {\cal J}}{\partial}^{-1}{\cal J})~'-
({\cal D}_{+}{\cal D}_{-}{\partial}^{-1}{\overline {\cal J}})^2\Bigr],
\label{eqs2jN=4}
\end{eqnarray}
\begin{eqnarray}
{\textstyle{\partial\over\partial t_3}} {\cal J} &=&
{\cal J}~''' +{\cal D}_{+}{\cal D}_{-}\Bigl\{3\Bigl[
{\cal J}~'{\partial}^{-1}{\overline {\cal J}} +
({\cal J}{\partial}^{-1}{\overline {\cal J}})
{\cal D}_{+} {\cal D}_{-}{\partial}^{-1}{\overline {\cal J}}
-\frac{1}{2}({\overline {\cal D}}^{+}{\overline {\cal D}}^{-}
{\partial}^{-1}{\cal J})^2\Bigr]~'\nonumber\\
&-&({\overline {\cal D}}^{+}{\overline {\cal D}}^{-}{\partial}^{-1}
{\cal J})^3 -3({\overline {\cal D}}^{+}{\overline {\cal D}}^{-}
{\partial}^{-1}{\cal J})^2
{\cal D}_{+}{\cal D}_{-}{\partial}^{-1}{\overline {\cal J}}+
6{\cal J}{\overline {\cal J}}
~{\overline {\cal D}}^{+} {\overline {\cal D}}^{-}{\partial}^{-1}
{\cal J}\Bigr\}, \nonumber\\
{\textstyle{\partial\over\partial t_3}}{\overline {\cal J}} &=&
{\overline {\cal J}}~'''  +
{\overline {\cal D}}^{+}{\overline {\cal D}}^{-}\Bigl\{3\Bigl[
-{\overline {\cal J}}~'{\partial}^{-1}{\cal J} +
({\overline {\cal J}}{\partial}^{-1}{\cal J})
{\overline {\cal D}}^{+} {\overline {\cal D}}^{-}{\partial}^{-1}{\cal J}
+\frac{1}{2}({\cal D}_{+}{\cal D}_{-}
{\partial}^{-1}{\overline {\cal J}})^2\Bigr]~'\nonumber\\
&-&({\cal D}_{+}{\cal D}_{-}{\partial}^{-1}
{\overline {\cal J}})^3 -3
({\cal D}_{+}{\cal D}_{-}{\partial}^{-1}{\overline {\cal J}})^2
{\overline {\cal D}}^{+}{\overline {\cal D}}^{-} {\partial}^{-1}{\cal J}+
6{\cal J}{\overline {\cal J}}
~{\cal D}_{+}{\cal D}_{-}{\partial}^{-1}{\overline {\cal J}}\Bigr\}
\label{eqs2jN=4t3}
\end{eqnarray}
and
\begin{eqnarray}
&&{\textstyle{\partial\over\partial t_2}} \Psi =
+{\Psi}~'' +2{\cal D}_-{\overline {\cal D}}^-
({\overline \Psi}~{\overline {\cal D}}^{-}\Psi)-
{\overline {\cal D}}^{+}({\cal D}_+\Psi)^2, \nonumber\\
&&{\textstyle{\partial\over\partial t_2}} {\overline \Psi} =
-{\overline \Psi}~'' - 2{\overline {\cal D}}^+{\cal D}_+
({\Psi}~{\cal D}_{+}{\overline \Psi})+
{\cal D}_{-}({\overline {\cal D}}^-{\overline \Psi})^2,
\label{eqslocal}
\end{eqnarray}
\begin{eqnarray}
{\textstyle{\partial\over\partial t_3}} \Psi  &=&
{\Psi}~''' +3{\cal D}_-
\Bigl[({\overline {\cal D}}^-{\Psi})~'
~{\overline {\cal D}}^{-}{\overline \Psi}
+({\overline {\cal D}}^{-}{\Psi})({\overline {\cal D}}^{-}
{\overline \Psi})^2+
\frac{1}{2}{\overline {\cal D}}^{+}{\overline {\cal D}}^{-}
({\cal D}_{+}{\Psi})^{2}\Bigr] \nonumber\\
&+&{\overline {\cal D}}^{+}\Bigl[
({\cal D}_{+}{\Psi})^{3}-
3({\cal D}_{+}{\Psi})^{2}
{\overline {\cal D}}^{-}{\overline \Psi}-
6({\cal D}_{+}{\overline \Psi})
({\overline {\cal D}}^{-}{\Psi})
{\cal D}_{+}{\Psi}\Bigr], \nonumber\\
{\textstyle{\partial\over\partial t_3}} {\overline \Psi}  &=&
{\overline {\Psi}}~''' +3{\overline {\cal D}}^+
\Bigl[({\cal D}_+{\overline {\Psi}})~'~{\cal D}_{+}{\Psi}
+({\cal D}_{+}{\overline {\Psi}})({\cal D}_{+}{\Psi})^2-
\frac{1}{2}{\cal D}_{+}{\cal D}_{-}
({\overline {\cal D}}^{-}{\overline {\Psi}})^{2}\Bigr] \nonumber\\
&+&{\cal D}_{-}\Bigl[
({\overline {\cal D}}^{-}{\overline {\Psi}})^{3}-
3({\overline {\cal D}}^{-}{\overline \Psi})^{2}
{\cal D}_{+}{\Psi}-
6({\cal D}_{+}{\overline \Psi})
({\overline {\cal D}}^{-}{\Psi})
{\overline {\cal D}}^{-}{\overline {\Psi}}\Bigr],
\label{eqslocalt3}
\end{eqnarray}
respectively.

\section{Hamiltonian structure of the N=4 Toda (KdV)
hierarchy in N=4 superspace}

Now, we would like to discuss the Hamiltonian structure
of the $N=4$ Toda (KdV) hierarchy in $N=4$ superspace.
Let us present first the general formulae
for the conserved quantities (Hamiltonians) $H^{t}_l$ of the $N=4$ flows
\p{flowsform} and (\ref{basisnew4}--\ref{b4}) in $N=4$ superspace,
\begin{eqnarray}
H^{t}_l = \int d z d \theta^+ d \eta^+ d \theta^- d \eta^-
{\partial}^{-1} \Big({({{\widetilde L}_1}^{T})}^{2l}\Big)_0.
\label{hamiltonians}
\end{eqnarray}
Hereafter, we use the following definitions of the $N=2$ and $N=4$
superspace integrals for an arbitrary superfield functional
$f(\theta^+ , \eta^+ , \theta^- ,\eta^- )$:
\begin{eqnarray}
\int d z d \theta^+ d \eta^+ d \theta^- d \eta^- f
(\theta^+ , \eta^+ , \theta^- ,\eta^- )  & \equiv &
\int d z d \theta^+ d \eta^+
\Big({\cal D}_{-}{\overline {\cal D}}^{-}
f\Big)\Big|_{\theta^{-}=\eta^{-}=0} \nonumber\\
& \equiv & \int d z \Big(
{\cal D}_{+}{\overline {\cal D}}^{+}
{\cal D}_{-}{\overline {\cal D}}^{-}
f\Big)\Big|_{\theta^{\pm}=\eta^{\pm}=0},
\label{integral}
\end{eqnarray}
respectively, as well as the following realization of the inverse
derivative:
\begin{eqnarray}
{\partial}_{z}^{-1} \equiv \frac{1}{2}
\int_{-\infty}^{+\infty}dx{\epsilon}(z-x), \quad
{\epsilon}(z-x)=-{\epsilon}(x-z)\equiv 1, \quad if \quad z>x.
\label{derrealiz}
\end{eqnarray}
Using these definitions
and the identities (\ref{relat2}--\ref{relat3}), one can equivalently
rewrite $H^{t}_l$ \p{hamiltonians} in the form of an $N=2$
superfield integral
\begin{eqnarray}
H^{t}_l = \int d z d \theta^+ d \eta^+
\Big({({{\widetilde L}_1}^{T})}^{2l}\Big)_0 \Big|_{\theta^{-}=\eta^{-}=0}.
\label{hamiltonians1}
\end{eqnarray}
Then, using the identity \p{eq1} as well as the relation
\p{laxgauge1}, one can easily show that this  representation
reproduces the $N=2$ superspace representation for the
Hamiltonians used in \cite{dgs}. Remembering that the quantities
$\Big({({{\widetilde L}_1}^{T})}^{2l}\Big)_0$ are local for
all the local flows (\ref{basisnew4}--\ref{b4}) (see the discussion after
equations \p{b5}), we are led to the conclusion that the corresponding
Hamiltonians \p{hamiltonians1} and their $N=2$ superfield densities are
local quantities as well, while the Hamiltonian densities in $N=4$
superspace entering into \p{hamiltonians} are nonlocal even when the flows
corresponding to them are local. It is interesting to remark that
$H^{t}_l$ \p{hamiltonians} can identically be rewritten in the following
form:
\begin{eqnarray}
H^{t}_l & = & {(-1)}^l \int d z d \eta^+ d \theta^+ d \theta^- d \eta^-
{\partial}^{-1} {\textstyle{\partial\over\partial t_l}} \xi \nonumber\\
& \equiv &
{(-1)}^l \int d z d \theta^+ d \eta^+ d \theta^- d \eta^-
({\cal D}_- + {\overline {\cal D}}^{-})^{-1}
({\cal D}_+ + {\overline {\cal D}}^{+})^{-1}
{\partial}^{-1} {\textstyle{\partial\over\partial t_l}}
({\overline {\cal J}}+{\cal J})],
\label{hamiltonians2}
\end{eqnarray}
where the second of equations \p{flowsg} and equation \p{definition} have been used.
We have verified that the following formula\footnote{The derivation of
\p{variation} makes essential use of the realization
\p{derrealiz} of the inverse derivative.}:
\begin{eqnarray}
\int d z d \theta^+ d \eta^+ d \theta^- d \eta^-
{\partial}^{-1} \frac{{\delta}}{{\delta} {\cal J}}
\Big({({{\widetilde L}_1}^{T})}^{2l}\Big)_0 =
l~{\overline {\cal D}}^{+}{\overline {\cal D}}^{-}{\partial}^{-2}
\Big({({{\widetilde L}_1}^{T})}^{2(l-1)}\Big)_0
\label{variation}
\end{eqnarray}
is valid for a few first values of $l$. The variation formula with
respect to ${\overline {\cal J}}$ can be obtained if one applies
the substitution \p{prop6} to equation \p{variation} and uses the identity
\p{prop8},
\begin{eqnarray}
\int d z d \theta^+ d \eta^+ d \theta^- d \eta^-
{\partial}^{-1} \frac{{\delta}}{{\delta} {\overline {\cal J}}}
\Big({({{\widetilde L}_1}^{T})}^{2l}\Big)_0 =
-l~{\cal D}_{+}{\cal D}_{-}{\partial}^{-2}
\Big({({{\widetilde L}_1}^{T})}^{2(l-1)}\Big)_0.
\label{variation1}
\end{eqnarray}
It is plausible to suppose that the formulae
(\ref{variation}--\ref{variation1}) are valid for all values of $l$ as
well, but we cannot present a proof here. Using these formulae
and the Hamiltonians $H^{t}_l$ \p{hamiltonians}, one can represent the
flows \p{flowsform} in the Hamiltonian form
\begin{eqnarray}
(-1)^{l+1}{\textstyle{\partial\over\partial t_{l-1}}}
\left(\begin{array}{cc} {\overline {\cal J}} \\ {\cal J}
\end{array}\right) = \frac{1}{l}
J_1 \left(\begin{array}{cc} {\delta}/{\delta {\overline {\cal J}}} \\
{\delta}/{\delta {\cal J}} \end{array}\right) H^{t}_{l}
= \frac{1}{l-1}
J_2 \left(\begin{array}{cc} {\delta}/{\delta {\overline {\cal J}}} \\
{\delta}/{\delta {\cal J}} \end{array}\right) H^{t}_{l-1},
\label{hameq}
\end{eqnarray}
where $J_1$,
\begin{eqnarray}
J_1 = \left(\begin{array}{cc} 0 &
-{\overline {\cal D}}^{+}{\overline {\cal D}}^{-}
{\cal D}_{+}{\cal D}_{-} \\
{\cal D}_{+}{\cal D}_{-}
{\overline {\cal D}}^{+}{\overline {\cal D}}^{-}&  0\end{array}\right),
\label{hameq1}
\end{eqnarray}
is the first Hamiltonian structure in $N=4$ superspace.
Using the flows (\ref{eqs2jN=4}--\ref{eqs2jN=4t3}), we have also
found the second Hamiltonian structure $J_2$
\begin{eqnarray}
&& \quad \quad \quad \quad \quad \quad \quad
J_2 = \left(\begin{array}{cc}
J_{11}& J_{12}\\J_{21}&  J_{22}\end{array}\right),\nonumber\\
&&J_{11}= -({\partial} {\overline {\cal J}}+
{\overline {\cal J}}{\partial})
{\overline {\cal D}}^{+}{\overline {\cal D}}^{-}, \quad
J_{22}=({\partial} {\cal J}+{\cal J}{\partial})
{\cal D}_{+}{\cal D}_{-}, \nonumber\\
&&J_{12}={\overline {\cal D}}^{+}{\overline {\cal D}}^{-}
\Big({\partial} +[{\cal D}_- {\cal D}_+ {\partial}^{-1}
{\overline {\cal J}}] + [{\overline {\cal D}}^{-}
{\overline {\cal D}}^{+}{\partial}^{-1}{\cal J}]\Big)
{\cal D}_{+}{\cal D}_{-}, \nonumber\\
&&J_{21}={\cal D}_{+}{\cal D}_{-}\Big({\partial} -
[{\cal D}_- {\cal D}_+ {\partial}^{-1}
{\overline {\cal J}}] - [{\overline {\cal D}}^{-}
{\overline {\cal D}}^{+}{\partial}^{-1}{\cal J}]\Big)
{\overline {\cal D}}^{+}{\overline {\cal D}}^{-}
\label{hameq2}
\end{eqnarray}
which is isomorphic to the $N=4$ SU(2) superconformal algebra (see the
end of this section and the remark at the end of section 5).
In terms of these two Hamiltonian structures the Poisson brackets of the
superfields ${\overline {\cal J}}$ and ${\cal J}$ are given by the
formula
\begin{eqnarray}
\{\left(\begin{array}{cc}
{\overline {\cal J}}(Z_1)\\ {\cal J}(Z_1)\end{array}\right)
\stackrel{\otimes}{,}
\left(\begin{array}{cc} {\overline {\cal J}}(Z_2),
{\cal J}(Z_2)\end{array}\right)\}_k=
J_k(Z_1){\delta}^{N=4}(Z_1-Z_2)
%, \quad {\delta}^{N=4}(Z)
%\equiv \delta(z) \theta^+ \eta^+ \theta^- \eta^-,
\label{palg}
\end{eqnarray}
where ${\delta}^{N=4}(Z)\equiv \delta(z) \theta^+ \eta^+ \theta^- \eta^-$
is the delta function in $N=4$ superspace with the coordinates
$Z \equiv \{z,\theta^+, \eta^+, \theta^-,$ $\eta^-\}$.
Let us  remark that the second Hamiltonian structure
$J_2$ \p{hameq2} can be rewritten in terms of the
superfunction $\xi$ \p{definition}, and it has the
following simple form:
\begin{eqnarray}
J_2 = \left(\begin{array}{cc}
-{\overline {\cal D}}^{+}{\overline {\cal D}}^{-}
({\partial}{\xi}+{\xi}{\partial})
{\overline {\cal D}}^{+}{\overline {\cal D}}^{-}&
{\overline {\cal D}}^{+}{\overline {\cal D}}^{-}
({\partial} +{\xi}~'){\cal D}_{+}{\cal D}_{-}\\
{\cal D}_{+}{\cal D}_{-}({\partial}-{\xi}~')
{\overline {\cal D}}^{+}{\overline {\cal D}}^{-}&
~{\cal D}_{+}{\cal D}_{-}({\partial} {\xi}+{\xi}{\partial})
{\cal D}_{+}{\cal D}_{-}\end{array}\right).
\label{hameq22}
\end{eqnarray}

Knowing the first and second Hamiltonian structures, we may
construct the recursion operator of the hierarchy in $N=4$ superspace
using the following general rule:
\begin{eqnarray}
R = J_2 {\widehat J}_1 \equiv
\left(\begin{array}{cc}
J_{12}, & -J_{11} \\
J_{22}, &  -J_{21}
\end{array}\right){\partial}^{-2}, \quad
{\textstyle{\partial\over\partial t_{l+1}}}
\left(\begin{array}{cc}
{\overline {\cal J}}\\{\cal J} \end{array}\right) =
R {\textstyle{\partial\over\partial t_{l}}}
\left(\begin{array}{cc} {\overline {\cal J}}\\
{\cal J} \end{array}\right), \quad
J_{l+1} = R^l J_1,
\label{recop0}
\end{eqnarray}
where the matrix ${\widehat J}_1$ is
\begin{eqnarray}
{\widehat J} =\left(\begin{array}{cc}
0, & -1 \\
1, &  0 \end{array}\right){\partial}^{-2}, \quad
{\widehat J}_1 J_1 \left(\begin{array}{cc}
{\overline {\cal J}}
\\{\cal J} \end{array}\right) =
J_1{\widehat J}_1  \left(\begin{array}{cc}
{\overline {\cal J}}
\\{\cal J} \end{array}\right) =
\left(\begin{array}{cc}
{\overline {\cal J}}
\\{\cal J} \end{array}\right).
\label{matrix-1}
\end{eqnarray}
The Hamiltonian structures $J_1$ and $J_2$ (\ref{hameq1}--\ref{hameq2})
are obviously compatible, e.g., the deformation
$[{\cal D}_- {\cal D}_+ {\partial}^{-1} {\overline {\cal J}}]$
$\Rightarrow$  $[{\cal D}_- {\cal D}_+ {\partial}^{-1}
{\overline {\cal J}}] +\alpha$,
where $\alpha$ is an arbitrary parameter, transforms $J_2$ into
the algebraic sum $J_2-\alpha J_1$. Thus, one concludes that the
recursion operator $R$ \p{recop0} is hereditary, being obtained
from a compatible pair of Hamiltonian structures \cite{ff}.

Applying formulae \p{recop0} we obtain the following recurrence relations
for the flows \p{flowsform} in $N=4$ superspace:
%%%%%%%%%%%%%%%%%%%%
%\begin{eqnarray}
%&&{\textstyle{\partial\over\partial t_{l+1}}}{\overline {\cal J}}=
%{\overline {\cal D}}^{+}{\overline {\cal D}}^{-}\Big(
%(+{\partial} +{\xi}~'){\cal D}_{+}{\cal D}_{-}+
%({\partial}{\xi}+{\xi}{\partial})
%{\overline {\cal D}}^{+}{\overline {\cal D}}^{-}\Big)
%{\partial}^{-2}{\textstyle{\partial\over\partial t_{l}}}
%({\overline {\cal J}}+{\cal J}), \nonumber\\
%&&{\textstyle{\partial\over\partial t_{l+1}}}{\cal J}=
%{\cal D}_{+}{\cal D}_{-}\Big((-{\partial}+{\xi}~')
%{\overline {\cal D}}^{+}{\overline {\cal D}}^{-}+
%({\partial} {\xi}+{\xi}{\partial})
%{\cal D}_{+}{\cal D}_{-}\Big)
%{\partial}^{-2}{\textstyle{\partial\over\partial t_{l}}}
%({\overline {\cal J}}+{\cal J}).
%\label{recrel}
%\end{eqnarray}
\begin{eqnarray}
&&{\textstyle{\partial\over\partial t_{l+1}}}{\overline {\cal J}}=
{\overline {\cal D}}^{+}{\overline {\cal D}}^{-}\Big(
(+{\partial} +{\xi}~'){\cal D}_{+}{\cal D}_{-}
{\partial}^{-2}{\textstyle{\partial\over\partial t_{l}}}
{\overline {\cal J}}+
({\partial}{\xi}+{\xi}{\partial})
{\overline {\cal D}}^{+}{\overline {\cal D}}^{-}
{\partial}^{-2}{\textstyle{\partial\over\partial t_{l}}}{\cal J}\Big),
\nonumber\\ &&{\textstyle{\partial\over\partial t_{l+1}}}{\cal J}=
{\cal D}_{+}{\cal D}_{-}\Big((-{\partial}+{\xi}~')
{\overline {\cal D}}^{+}{\overline {\cal D}}^{-}
{\partial}^{-2}{\textstyle{\partial\over\partial t_{l}}}
{\cal J} + ({\partial} {\xi}+{\xi}{\partial})
{\cal D}_{+}{\cal D}_{-}
{\partial}^{-2}{\textstyle{\partial\over\partial t_{l}}}
{\overline {\cal J}}\Big).
\label{recrel}
\end{eqnarray}
These relations can be rewritten in terms of the {\it single}
dimensionless superfunction ${\xi}$ \p{definition} in the following
form:
%%%%%%%%%%%%%%%%%%%%%%%%%%%%%%%
%\begin{eqnarray}
%{\textstyle{\partial\over\partial t_{l+1}}}{\xi}&=&
%-{\partial}^{-2}\Big\{{\cal D}_{+}{\cal D}_{-}
%{\overline {\cal D}}^{+}{\overline {\cal D}}^{-}
%\Big((+{\partial} +{\xi}~'){\cal D}_{+}{\cal D}_{-}
%{\overline {\cal D}}^{+}{\overline {\cal D}}^{-}+
%({\partial}{\xi}+{\xi}{\partial})
%{\overline {\cal D}}^{+}{\overline {\cal D}}^{-}
%{\cal D}_{+}{\cal D}_{-}\Big)
%\nonumber\\ &+&{\overline {\cal D}}^{+}{\overline {\cal D}}^{-}
%{\cal D}_{+}{\cal D}_{-}\Big((-{\partial}+{\xi}~')
%{\overline {\cal D}}^{+}{\overline {\cal D}}^{-}
%{\cal D}_{+}{\cal D}_{-}+ ({\partial} {\xi}+{\xi}{\partial})
%{\cal D}_{+}{\cal D}_{-}
%{\overline {\cal D}}^{+}{\overline {\cal D}}^{-}\Big)
%\Big\}{\partial}^{-2}{\textstyle{\partial\over\partial t_{l}}}{\xi}.
%\label{recrelgreat1}
%\end{eqnarray}
%%%%%%%%%%%%%%%%%%%%%%%%%%%%%%%%%%%
%\begin{eqnarray}
%{\textstyle{\partial\over\partial t_{l+1}}}{\xi}=
%-\Big\{{\Pi}\Big((+{\partial} +{\xi}~'){\Pi}+
%({\partial}{\xi}+{\xi}{\partial}){\overline {\Pi}}\Big)
%+ {\overline {\Pi}}\Big((-{\partial}+{\xi}~')
%{\overline {\Pi}}+ ({\partial} {\xi}+{\xi}{\partial}){\Pi}\Big)
%\Big\}{\textstyle{\partial\over\partial t_{l}}}{\xi}
%\label{recrelgreat1}
%\end{eqnarray}
%%%%%%%%%%%%%%%%%%%%%%%%%%%%%%%%%%%%%%%%%%%
\begin{eqnarray}
{\textstyle{\partial\over\partial t_{l+1}}}{\xi}=
\Big(({\overline {\Pi}}-{\Pi}){\partial}-
({\Pi}+{\overline {\Pi}}){\xi}~'
%({\Pi}+{\overline {\Pi}})
-2{\Pi}{\xi}{\overline {\Pi}}{\partial}-
2{\overline {\Pi}}{\xi}{\Pi}{\partial}\Big)
{\textstyle{\partial\over\partial t_{l}}}{\xi},
\label{recrelgreat}
\end{eqnarray}
where the operators ${\Pi}$ and ${\overline {\Pi}}$ are $N=4$ chiral
projectors
\begin{eqnarray}
&& \quad {\Pi}\equiv -{\cal D}_{+}{\cal D}_{-} {\overline {\cal D}}^{+}
{\overline {\cal D}}^{-}{\partial}^{-2},\quad
{\overline {\Pi}}\equiv -{\overline {\cal D}}^{+}{\overline {\cal D}}^{-}
{\cal D}_{+}{\cal D}_{-}{\partial}^{-2},  \nonumber\\ &&
{\Pi}^{2}={\Pi}, \quad {\overline {\Pi}}^{2}={\overline {\Pi}}, \quad
{\Pi}{\overline {\Pi}}={\overline {\Pi}}{\Pi}=0, \quad
({\Pi}+{\overline {\Pi}}){\xi}={\xi}.
\label{defproj}
\end{eqnarray}
The superfunction ${\xi}$ can actually be treated as an independent
constrained real linear $N=4$ superfield \cite{ggrs}
\begin{eqnarray}
{\cal D}_{+}{\overline {\cal D}}^{-}{\xi}=0, \quad
{\cal D}_{-}{\overline {\cal D}}^{+}{\xi}=0
\label{constrxi}
\end{eqnarray}
which we could have taken as our starting point,
while the relation \p{definition} in this case
is the general solution of the quadratic constraints \p{constrxi}
expressed in terms of the pair of chiral-antichiral superfields ${\cal J}$
and ${\overline {\cal J}}$ \p{N=4constr}. Conversely, one can
express ${\cal J}$ and ${\overline {\cal J}}$ in terms of $\xi$
\begin{eqnarray}
{\cal J}={\cal D}_+{\cal D}_{-}{\xi}, \quad
{\overline {\cal J}}={\overline {\cal D}}^{+}
{\overline {\cal D}}^{-}{\xi}.
\label{j-xi}
\end{eqnarray}
Let us also remark that the last relation in equations \p{defproj} is
a direct consequence of the constraints \p{constrxi} too. Therefore, we
come to the conclusion that all the flows of the N=4 Toda (KdV)
hierarchy are actually the flows for a single real linear $N=4$
superfield ${\xi}$ \p{constrxi}. Moreover, these flows are local because
the Lax operator ${{\widetilde L}_1}^{T}$ \p{laxrel2g1new} entering into
the flow equations \p{flowsg} for the superfield ${\xi}$, being expressed
in terms of ${\xi}$ via \p{j-xi}
\begin{eqnarray}
{{\widetilde L}_1}^T \equiv
-{\cal D}_- - {\overline {\cal D}}^{-} +
[({\cal D}_- + {\overline {\cal D}}^{-}){\xi}]-
{({\cal D}_+ +{\overline {\cal D}}^{+})}^{-1}~
[{\overline {\cal D}}^{+}{\overline {\cal D}}^{-}{\xi}],
\label{laxrel2g1newnew}
\end{eqnarray}
is local in ${\xi}$. As an illustrative example, let us present
explicitly the second flow
\begin{eqnarray}
{\textstyle{\partial\over\partial t_2}} {\xi} =
({\overline {\cal D}}^{+}{\cal D}_{+}-
{\cal D}_{-}{\overline {\cal D}}^{-})
{\xi}~' -({\xi}~')^{2} +
2({\cal D}_{+}{\cal D}_{-}{\xi})
({\overline {\cal D}}^{+}{\overline {\cal D}}^{-}{\xi})
\label{eqs2jN=4xi}
\end{eqnarray}
resulting from the recurrence relation \p{recrelgreat} and the first flow
${\textstyle{\partial\over\partial t_{1}}}{\xi}={\xi}~'$.
In the derivation of the flows \p{eqs2jN=4xi}, one has to use the identity
\begin{eqnarray}
\quad \quad {\cal D}_{+}{\overline {\cal D}}^{+}{\xi}=
{\cal D}_{-}{\overline {\cal D}}^{-}{\xi}, \quad
\label{constrxi-ident}
\end{eqnarray}
following from the constraints \p{constrxi}.
For completeness, we also present the first
\begin{eqnarray}
&& ~~\{{\xi}(Z_1),{\xi}(Z_2)\}_{k}=J^{\xi}_k(Z_1){\delta}^{N=4}(Z_1-Z_2),
\nonumber\\&& J^{\xi}_1={\widehat J}^{\xi}_1=
({\overline {\Pi}}-{\Pi}), \quad
J^{\xi}_1{\widehat J}^{\xi}_1{\xi}=
{\widehat J}^{\xi}_1J^{\xi}_1{\xi}={\xi}
\label{hamstrxi1}
\end{eqnarray}
and second
\begin{eqnarray}
J^{\xi}_2 &=&
({\overline {\Pi}}+{\Pi}){\partial}+
({\Pi}+{\overline {\Pi}}){\xi}~'({\Pi}-{\overline {\Pi}})-
2{\Pi}{\xi}{\overline {\Pi}}{\partial}+
2{\overline {\Pi}}{\xi}{\Pi}{\partial}\nonumber\\
&\equiv & {\partial}^{-1}\Big(
({\Pi}-{\overline {\Pi}}){\partial}^{2}+{\partial}{\xi}'
-({\cal D}_{+}{\xi}'){\overline {\cal D}}^{+}
-({\overline {\cal D}}^{+}{\xi}'){\cal D}_{+}
-({\cal D}_{-}{\xi}'){\overline {\cal D}}^{-}
-({\overline {\cal D}}^{-}{\xi}'){\cal D}_{-}
\nonumber\\ &-&2({\cal D}_{+}{\cal D}_{-}{\xi})
{\overline {\cal D}}^{+}{\overline {\cal D}}^{-}-
2({\overline {\cal D}}^{+}{\overline {\cal D}}^{-}{\xi})
{\cal D}_{+}{\cal D}_{-}
\Big)({\Pi}-{\overline {\Pi}})
\label{hamstrxi2}
\end{eqnarray}
Hamiltonian structures as well as the recursion operator
\begin{eqnarray}
R^{\xi} = J^{\xi}_2 {\widehat J}^{\xi}_1=
({\overline {\Pi}}-{\Pi}){\partial}-
({\Pi}+{\overline {\Pi}}){\xi}~'({\Pi}+{\overline {\Pi}})-
2{\Pi}{\xi}{\overline {\Pi}}{\partial}-
2{\overline {\Pi}}{\xi}{\Pi}{\partial}
\label{recopxi}
\end{eqnarray}
of the $N=4$ Toda (KdV) hierarchy described in terms of the
superfield ${\xi}$ \p{constrxi}.

%zdes'
To close this section we would like to note that
four closed subalgebras, isomorphic to the $N=4$ $SU(2)$
superconformal algebra (SCA), can be
extracted from the $N=4$ superfield form
of the $N=4$ $O(4)$ (large) SCA,
\cite{sch} (see also references
therein)\footnote{Some $N=4$ $SU(2)$ SC subalgebras of the $N=4$ $O(4)$ SCA
were discussed in \cite{sch,stp} at the component level.}
\begin{eqnarray}
\{{\sigma},{\sigma}\}=
\Big(c_1{\partial}+c_2 ({\Pi}+{\overline {\Pi}}){\partial}
+ {\sigma}' -({\cal D}_{+}{\sigma}){\overline {\cal D}}^{+}
-({\overline {\cal D}}^{+}{\sigma}){\cal D}_{+}
-({\cal D}_{-}{\sigma}){\overline {\cal D}}^{-}
-({\overline {\cal D}}^{-}{\sigma}){\cal D}_{-}\Big){\delta}^{N=4},
\label{N=4large}
\end{eqnarray}
where ${\sigma}={\sigma}(Z)$ is the $N=4$ $O(4)$ unconstrained
supercurrent with $16$
independent field components. $c_1$ and $c_2$ are two independent
central charges. These subalgebras are defined using the operators
\begin{eqnarray}
&&{\pi}_{\pm}\equiv ({\cal D}_{+}{\cal D}_{-}\pm {\overline {\cal D}}^{+}
{\overline {\cal D}}^{-}){\partial}^{-1}, \quad
{\cal D}_{+}{\overline {\cal D}}^{-}{\pi}_{\pm}=
{\cal D}_{-}{\overline {\cal D}}^{+}{\pi}_{\pm}=
{\pi}_{\pm}{\cal D}_{+}{\overline {\cal D}}^{-}=
{\pi}_{\pm}{\cal D}_{-}{\overline {\cal D}}^{+}=0, \nonumber\\
&& {\pi}_{\pm}^{2n}= (\mp 1)^{n}
({\Pi}+{\overline {\Pi}}), \quad
{\pi}_{\pm}^{2n+1}= (\mp 1)^{n}{\pi}_{\pm}, \quad
{\pi}_{+}{\pi}_{-}=({\Pi}-{\overline {\Pi}})
\quad \{{\pi}_{+},{\pi}_{-}\}=0 ,
%zdes'
\label{pi}
\end{eqnarray}
where the operators $\{{\pi}_{+}, {\pi}_{-}, {\pi}_{+}^{2},
{\pi}_{+}{\pi}_{-}\}$  form a linearly independent set of degenerate
operators satisfying the constraint \p{constrxi}.
Then we may define four
supercurrents expressed in terms of ${\sigma}$ by
\begin{eqnarray}
\xi_{\pm} = {\pi}_{\pm}{\sigma}, \quad {\widehat {\xi}} =
{\pi}_{+}^{2}{\sigma}, \quad  \xi = {\pi}_{+}{\pi}_{-}{\sigma}.
\label{relations}
\end{eqnarray}
Each of them satisfies the constraints \p{constrxi} and, as a
consequence, possesses only $8$ independent superfield components.
The degenerate mapping
${\xi}$ to ${\sigma}$
\p{relations} gives the precise relationship between the Poisson
structures \p{hamstrxi2} and \p{N=4large}. The central
charge of the $N=4$ $SU(2)$ SC subalgebra of the $N=4$ $O(4)$ SCA is
$c=c_1+c_2$. The superfield relationship \p{relations}
between the $N=4$ superconformal $SU(2)$ and $O(4)$ algebras could give a
bridge towards the resolution of the longstanding problem of constructing an
$N=4$ $O(4)$ KdV hierarchy (if any), but this problem is out of the scope
of the present paper and will be considered elsewere.

\section{Real forms and discrete symmetries of the N=4 Toda (KdV)
hierarchy in N=4 superspace}
It is well known that different real forms derived from the same complex
integrable hierarchy are nonequivalent in general.
Keeping this in mind it seems important to find as many
different real forms of the $N=4$ Toda (KdV) hierarchy as possible.
This task was already analysed in \cite{ik,ds} at the level of the second
and third flows of the hierarchy in the real $N=2$ superspace, and
three nonequivalent complex conjugations were constructed for
these flows. Nonequivalent here means that
it is not possible to relate any two of them  via obvious symmetries.
Now, based on the results of the previous sections
we are able to extend these complex conjugations to $N=4$ superspace
and prove that all the flows ${\textstyle{\partial\over\partial t_l}}$
of the hierarchy in $N=4$ superspace are invariant under any of
these complex conjugations.

We will use the standard convention regarding complex
conjugation of products involving odd operators and functions
(see, e.g., the books \cite{ggrs,w}). In particular, if ${\hbox{\bbd O}}$
is some even differential operator acting on a superfield $F$, we define
the complex conjugate of ${\hbox{\bbd O}}$ by $({\hbox{\bbd O}}F)^*=
{\hbox{\bbd O}}^*F^*$.

Let us prove the statement

{\it the flows \p{flowsform} are invariant under the following three complex
conjugations}:
\begin{eqnarray}
({\cal J},~{\overline {\cal J}})^{*}=-({\cal J},~{\overline {\cal J}}),
\quad (z,{{\theta}^{\pm}},\eta^{\pm})^{*}=(-z,{\theta}^{\pm},-\eta^{\pm}),
\quad t^{*}_l=(-1)^{l}t_l,
\label{conj1}
\end{eqnarray}
\begin{eqnarray}
({\cal J},~{\overline {\cal J}})^{\bullet}=
(~{\cal J}- {\cal D}_-{\cal D}_+\ln
{\overline {\cal J}},~{\overline {\cal J}}~),
\quad (z,{{\theta}^{\pm}},\eta^{\pm})^{\bullet}=
(-z,{\theta}^{\pm},-\eta^{\pm}), \quad t^{\bullet}_l=-t_l,
\label{conj2}
\end{eqnarray}
\begin{eqnarray}
({\cal J},~{\overline {\cal J}})^{\star}
=({\overline {\cal J}},~{\cal J}),
\quad (z,{{\theta}^{\pm}},\eta^{\pm})^{\star}=
(-z,{\theta}^{\pm},\eta^{\pm}), \quad t^{\star}_l=-t_l.
\label{conj3}
\end{eqnarray}
With this aim let us describe the
properties of the Lax operator ${\widetilde L}^{T}_1$ \p{laxrel2g1new}
and of the quantity $\Big({({{\widetilde L}_1}^{T})}^{2l}\Big)_0$
entering into the flow equations \p{flowsform} under these involutions. They are
\begin{eqnarray}
({\widetilde L}^{T}_1)^{*}={\widetilde L}^{T}_1, \quad
\Big({({{\widetilde L}_1}^{T})}^{2l}\Big)_0^{*}=
(-1)^{l}\Big({({{\widetilde L}_1}^{T})}^{2l}\Big)_0
\label{conj1lax}
\end{eqnarray}
\begin{eqnarray}
({\widetilde L}^{T}_1)^{\bullet}=-\frac{1}{{\overline {\cal J}}}
~{\widetilde L}_1~{\overline {\cal J}},
\quad  \Big({({{\widetilde L}_1}^{T})}^{2l}\Big)_0^{\bullet}=
\Big({({{\widetilde L}_1}^{T})}^{2l}\Big)_0 +
\frac{1}{{\overline {\cal J}}}
~[~{{\Big(\Big(({{\widetilde L}_1}^{T})}^{2l}
\Big)_{\ge 1}\Big)}^{T}~ {\overline {\cal J}}~],
\label{conj2lax}
\end{eqnarray}
\begin{eqnarray}
({\widetilde L}^{T}_1)^{\star}
= ({\cal D}_+ +{\overline {\cal D}}_+)^{-1} {\widetilde L}_1
({\cal D}_+ +{\overline {\cal D}}_+), \quad
\Big({({{\widetilde L}_1}^{T})}^{2l}\Big)_0^{\star}=
\Big({({{\widetilde L}_1}^{T})}^{2l}\Big)_0,
\label{conj3lax}
\end{eqnarray}
where the identities \p{identity} and \p{prop9}
have been used when deriving equations \p{conj2lax} and \p{conj3lax},
respectively. Now, using these relations as well as the identity
\p{relat1} it is a simple exercise to verify that the flows \p{flowsform}
are indeed invariant under the complex conjugations (\ref{conj1}--\ref{conj3}).

The complex conjugations (\ref{conj1}--\ref{conj3}) extract different real
forms of the algebra \p{algnn4}. It may be verified that the real forms of the algebra
\p{algnn4} with the involutions (\ref{conj1}--\ref{conj2}) correspond to a
twisted real $N=4$ supersymmetry, while the real form corresponding to the
involution \p{conj3} reproduces the algebra of the real $N=4$
supersymmetry.

We would like to close this section with a few remarks.

First, a combination of the two complex conjugations
\p{conj3} and \p{conj2} being applied twice generates a manifestly $N=4$
supersymmetric form of the $N=4$ Toda chain equations\footnote{This way
of deriving discrete symmetries was proposed in \cite{s} and applied to
the construction of discrete symmetry transformations of the $N=2$
supersymmetric GNLS hierarchies.}
\begin{eqnarray}
{\cal J}^{\star \bullet \star \bullet}=
{\cal J} -{\cal D}_-{\cal D}_+\ln {\overline {\cal J}}, \quad
{\overline {\cal J}}^{\star \bullet \star \bullet}={\overline {\cal J}} -
{\overline {\cal D}}^-{\overline {\cal D}}^+
\ln {\cal J}^{\star \bullet \star\bullet}
\label{conj2jn=4}
\end{eqnarray}
which is the one-dimensional reduction of the two-dimensional $N=(2|2)$
superconformal Toda lattice \cite{eh,lds}. They form an
infinite-dimensional group of discrete Darboux symmetries of the $N=4$
supersymmetric Toda chain (KdV) hierarchy \cite{ls,ols,dgs,ds}.
In other words, if the set $\{{\cal J},{\overline {\cal J}} \}$
is a solution of the $N=4$ Toda (KdV) hierarchy, then the set
$\{{\cal J}^{\star \bullet \star \bullet},
{\overline {\cal J}}^{\star \bullet \star \bullet}\}$, related to
the former by equations \p{conj2jn=4}, is a solution of the hierarchy
as well. Let us also present equations \p{conj2jn=4}
written in the $N=4$ superfield basis
$\{{\Omega}, {\overline {\Omega}}\}$ \p{basisnew1},
where the $N=4$ flows are local
\begin{eqnarray}
{\overline \Omega}^{\star \bullet \star \bullet}=
{\overline \Omega} + {\cal D}_-\ln [{\overline {\cal D}}^{+}\Omega], \quad
{\Omega}^{\star \bullet \star \bullet}={\Omega}+{\overline {\cal D}}^-
\ln [{\cal D}_+{\overline {\Omega}}^{\star \bullet \star\bullet}].
\label{newbasisjn=4}
\end{eqnarray}

Second, a combination of the two complex conjugations
\p{conj1} and \p{conj2} generates an order two discrete
symmetry of the odd flows ${\textstyle{\partial\over\partial t_{2l+1}}}$,
\begin{eqnarray}
({\cal J},~{\overline {\cal J}})^{\bullet *}=
-(~{\cal J} + {\cal D}_-{\cal D}_+\ln
{\overline {\cal J}},~{\overline {\cal J}}~), \quad
({\cal J},~{\overline {\cal J}})^{\bullet *\bullet *}=
({\cal J},~{\overline {\cal J}}).
\label{symm}
\end{eqnarray}

Third, a combination of the two complex conjugations \p{conj3} and
\p{conj1} generates the substitution \p{prop6}, $\# \equiv \star *$.

Fourth, a direct verification shows that the second and third flows
(\ref{eqs2jN=4}--\ref{eqs2jN=4t3}) as well as the recursion relations
\p{recrel} are invariant under two additional complex conjugations
\begin{eqnarray}
({\cal J},~{\overline {\cal J}})^{\dagger}=-({\cal J},
~{\overline {\cal J}}), \quad
(z,{{\theta}^{\pm}},\eta^{\pm})^{\dagger}=(-z,i{\eta}^{\pm},i\theta^{\pm}),
\quad t^{\dagger}_l=(-1)^{l}t_l,
\label{conj1new}
\end{eqnarray}
\begin{eqnarray}
({\cal J},~{\overline {\cal J}})^{\ddagger}=({\cal J},
~{\overline {\cal J}}), \quad
(z,{{\theta}^{\pm}},\eta^{\pm})^{\ddagger}=(-z,{\theta}^{\mp},-\eta^{\mp}),
\quad t^{\ddagger}_l=(-1)^{l}t_l.
\label{conj2new}
\end{eqnarray}
It means that all the flows \p{flowsform} of the hierarchy are invariant under these
complex conjugations as well.

\section{N=4 Toda (KdV) hierarchy in N=2 superspace}
Let us present the relationship of the formulation of the $N=4$ Toda (KdV)
hierarchy in the $N=4$ superspace developed in previous sections
with its description in the three different $N=2$ superfield bases (a),
(b) and (c) from \cite{dik} (see equations (4.5) and (4.3a,b,c) therein)
at the level of the second flow
equations \p{eqs2jN=4}. As a byproduct, a correspondence with
the formalism of references \cite{di,dik} will be established.

{}~

~~~~~~~\noindent{\bf The basis (a) $\{{\widetilde V}, F, {\overline F}\}$.}

\begin{eqnarray}
(a) \quad
&& {\widetilde V}\equiv \frac{1}{2}
\Big[D_{-}D_{+}^{-1}({\overline {\cal J}}
+{\cal J})\Big](z,\theta^+,\theta^-=0,\eta^+, \eta^-=0), \nonumber\\
&&F \equiv {\cal J}(z,\theta^+,\theta^-=0,\eta^+, \eta^-=0), \nonumber\\
&&{\overline F}\equiv {\overline {\cal J}}(z,\theta^+,\theta^-=0,\eta^+,
\eta^-=0),
\label{a}
\end{eqnarray}
where ${\widetilde V}\equiv {\widetilde V}(z,\theta^+,\eta^+)$,
$F\equiv F(z,\theta^+,\eta^+)$ and
${\overline F}\equiv {\overline F}(z,\theta^+,\eta^+)$
are new unconstrained, chiral and antichiral,
\begin{eqnarray}
{\cal D}_+ ~F=0, \quad {\overline {\cal D}}^{+}~{\overline F} =0,
\label{chiral1}
\end{eqnarray}
even $N=2$ superfields, respectively, and $D_{\pm}$ is defined in equation
\p{lax3}. Using the chirality constraints
\p{N=4constr} and the
definition of the superfield ${\widetilde V}$ \p{a} one can express the
derivatives ${\cal D}_{-}{\overline {\cal J}}$ and
${\overline {\cal D}}^{-}{\cal J}$ in terms of
${\cal D}_{+}{\widetilde V}$ and ${\overline {\cal D}}^{+}{\widetilde V}$
\begin{eqnarray}
&{\overline {\cal D}}^{-}{\cal J}= -2 {\cal D}_{+}{\widetilde V}, \quad
\quad {\cal D}_{-}{\overline {\cal J}} = -2{\overline {\cal D}_{+}}
{\widetilde V}.&
\label{N=4n2relexpra}
\end{eqnarray}
Using these inputs, one can transform \p{eqs2jN=4} to the
superfield basis $\{ {\widetilde V},F, {\overline F}\}$ where
they take the following local form:
\begin{eqnarray}
&&{\textstyle{\partial\over\partial t_2}} {\widetilde V} =
([{\cal D}_+,{\overline {\cal D}}^{+}]{\widetilde V}+2{\widetilde V}^2 -
F{\overline F})~', \nonumber\\
&&{\textstyle{\partial\over\partial t_2}}F = -F~''
+4{\cal D}_+{\overline {\cal D}}^{+}(F{\widetilde V}),\nonumber\\
&&{\textstyle{\partial\over\partial t_2}}{\overline F} = +{\overline F}~''
+4{\overline {\cal D}}^{+}{\cal D}_+({\overline F}{\widetilde V}).
\label{releqa}
\end{eqnarray}

{}~

~~~~~~~\noindent{\bf The basis (c) $\{ {\widetilde J},\Phi,
{\overline \Phi}\}$.}

\begin{eqnarray}
(c) \quad
&&\frac{1}{2}({\Phi}+{\overline {\Phi}})-i{\widetilde J}\equiv
{\cal J}(z,\theta^+,\theta^-,\eta^+=0, \eta^-=0), \nonumber\\
&&\frac{1}{2}({\Phi}+{\overline {\Phi}})+i{\widetilde J}\equiv
{\overline {\cal J}}(z,\theta^+,\theta^-,\eta^+=0, \eta^-=0),
\label{c}
\end{eqnarray}
\begin{eqnarray}
{\widetilde J} \equiv \frac{i}{2}({\cal J}-
{\overline {\cal J}})\Big|_{\eta_{\pm} =0}, \quad
\Phi \equiv \Big[D{\overline D}{\partial}^{-1}({\overline {\cal J}}+
{\cal J})\Big]\Big|_{\eta_{\pm} =0},\quad
{\overline \Phi} \equiv \Big[{\overline D}D{\partial}^{-1}
({\overline {\cal J}}+{\cal J})\Big]\Big|_{\eta_{\pm} =0},
\label{invc}
\end{eqnarray}
where ${\widetilde J}\equiv {\widetilde J}(z,\theta^+,\theta^-),
\Phi\equiv \Phi(z,\theta^+,\theta^-)$ and
${\overline \Phi}\equiv {\overline \Phi}(z,\theta^+,\theta^-)$
are new unconstrained, chiral and antichiral,
\begin{eqnarray}
D ~{\Phi}=0, \quad {\overline D}~ {\overline {\Phi}} =0,
\label{chiral2}
\end{eqnarray}
even $N=2$ superfields,
respectively, and $D, \overline D$ are $N=2$ odd covariant derivatives,
\begin{eqnarray}
D\equiv \frac{1}{2}(D_{+}+iD_{-}), \quad
{\overline D}\equiv \frac{1}{2}(D_{+}-iD_{-}), \quad
\{D,{\overline D}\}={\partial}, \quad D^2={\overline D}^2=0.
\label{relcovder}
\end{eqnarray}
Using the explicit realization of the odd derivatives
$D_{\pm}$ \p{alg} and ${\cal D}_{\pm}$, ${\overline {\cal D}}^{\pm}$
\p{algnn4} as well as the chirality constraints \p{N=4constr}, one can
express the derivatives ${\cal D}_{\pm}{\overline {\cal J}}$ and
${\overline {\cal D}}^{\pm}{\cal J}$ in terms of
$D_{\pm}{\overline {\cal J}}$ and $D_{\pm}{\cal J}$
\begin{eqnarray}
&&{\overline {\cal D}}^{\pm}{\cal J}= D_{\pm}{\cal J}, \quad
{\overline {\cal D}}^{+}~ {\overline {\cal D}}^{-}{\cal J} =
D_+D_{-}{\cal J}, \quad {\cal D}_{\pm}{\overline {\cal J}} =
D_{\pm}{\overline {\cal J}}, \quad
{\cal D}_{+}~{\cal D}_{-} {\overline {\cal J}} =
D_{+}D_{-}{\overline {\cal J}}.
\label{N=4n2relexpr}
\end{eqnarray}
Using these inputs, one can transform equations \p{eqs2jN=4} to the
superfield basis $\{ {\widetilde J},\Phi, {\overline \Phi}\}$ where
they take the following local form:
\begin{eqnarray}
&&-i{\textstyle{\partial\over\partial t_2}} {\widetilde J} =
-\frac{1}{2}(\Phi + {\overline \Phi})~'' -
2({\widetilde J} (\Phi - {\overline \Phi}))~'+
[D,{\overline D}]({\widetilde J} (\Phi + {\overline \Phi})), \nonumber\\
&&-i{\textstyle{\partial\over\partial t_2}}{\Phi} =
2D{\overline D}({\widetilde J}~' -{\widetilde J}^2
-\frac{3}{4} {\Phi}^2 +\frac{1}{2}{\Phi} {\overline \Phi}), \nonumber\\
&&-i{\textstyle{\partial\over\partial t_2}}{\overline \Phi} =
2{\overline D}D({\widetilde J}~' +{\widetilde J}^2
+\frac{3}{4} {\overline \Phi}^2 -\frac{1}{2}{\Phi} {\overline \Phi}).
\label{releqc}
\end{eqnarray}
It is also instructive to present explicitly the Lax operator $L$
\p{lax2} expressed in terms of the $N=2$ superfields
$\{ {\widetilde J},\Phi, {\overline \Phi}\}$,
\begin{eqnarray}
L \equiv i({\overline D} - D)-\frac{1}{2}({\Phi}+{\overline {\Phi}}+
2i{\widetilde J})~\frac{1}{{\overline D}+D+
i[{\partial}^{-1} ({\overline D}{\Phi}-D{\overline {\Phi}})]} \quad .
\label{lax2new}
\end{eqnarray}

{}~

~~~~~~~\noindent{\bf The basis (b).}

{}~

The $N=2$ superfield basis (b) and the corresponding second flow
equations can be obtained from equations (\ref{c}--\ref{invc}) and \p{releqc}
of the basis (c) by the following substitution:
\begin{eqnarray}
(b) \quad \{{\widetilde J},\Phi, {\overline \Phi}\} \Rightarrow
\{{\widetilde J},i\Phi, -i{\overline \Phi}\}, \quad t_2 \Rightarrow it_2.
\label{b}
\end{eqnarray}

Therefore, we are led to the conclusion that the second flow equations
\p{eqs2jN=4} unify the three particular "$SU(2)$ frames" of the $N=4$ KdV
hierarchy in $N=2$ superspace (compare equations \p{releqa} and \p{releqc} with
equations (4.5) and (4.3a,b,c) from \cite{dik}). The same property is
certainly valid for any other flow of the $N=4$ Toda (KdV) hierarchy in
the $N=4$ superspace.

It is interesting to remark that if one introduces an auxiliary $N=4$
superfield $V^{12}$ as
\begin{eqnarray}
V^{12} \equiv V^{21} \equiv
-\frac{i}{2}D_{-}D_{+}^{-1}({\overline {\cal J}}+{\cal J})
\label{def}
\end{eqnarray}
as well as the following notations:
\begin{eqnarray}
V^{11}\equiv -i~{\overline {\cal J}}, \quad V^{22}\equiv i~{\cal J}, \quad
{\cal D}_1 \equiv {\cal D}_{+}, \quad
{\cal D}_2 \equiv  {\overline {\cal D}}^{-}, \quad
{\overline {\cal D}}^1 \equiv  {\overline {\cal D}}^{+}, \quad
{\overline {\cal D}}^2 \equiv  {\cal D}_{-},
\label{def1}
\end{eqnarray}
then the constraints \p{N=4constr} and \p{def} can equivalently be
rewritten in the covariant form:
\begin{eqnarray}
{\cal D}^{(i}V^{jk)} = 0, \quad {\overline {\cal D}}^{(i}V^{jk)} = 0,
\quad i,j,k = 1,2,
\label{def2}
\end{eqnarray}
where the indices $i,j,k$ are raised and lowered by the antisymmetric
tensors ${\epsilon}^{ij}$ and ${\epsilon}_{ij}$,
respectively, (${\epsilon}^{ij}{\epsilon}_{jk}= {{\delta}^{i}}_k$,
${\epsilon}_{12}=-{\epsilon}^{12}=1$) and $(ijk)$ means symmetrization.
The $SU(2)$ spin 1 $N=4$ supercurrent $V^{ij}=V^{ji}$ was initially
introduced in \cite{di,dik} for the description of the $N=4$ KdV equation.
Thus, formulae (\ref{def}--\ref{def1}) establish the precise
correspondence\footnote{It is also necessary to replace the
coordinate $z$ by $iz$.}
%as well as the fermionic covariant derivatives
%(${\cal D}^{i}$) ${\overline {\cal D}^{i}}$ by
%$({\overline {\cal D}^{i}})$ ${\cal D}^{i}$.}
with \cite{di,dik}. Using this correspondence, one
can calculate, e.g. the matrices $a^{ij}$ from \cite{dik},
\begin{eqnarray}
a=\left(\begin{array}{cc} 0 & 1 \\ 1 & 0 \end{array}\right), \quad
a=\left(\begin{array}{cc} 1 & ~~0 \\ 0 & -1 \end{array}\right)
\label{matricesa}
\end{eqnarray}
which correspond to \p{releqa} and \p{releqc}, respectively, as well as
establish the one--to--one correspondence of the second Hamiltonian
structure $J_2$ \p{hameq2} with the $N=4$ superspace form of the $N=4$
SU(2) superconformal algebra used in \cite{dik}.

\section{Alternative Lax pair construction in N=4 superspace}

The results of the previous sections can be used as inputs for an alternative,
heuristic construction of a Lax pair representation of the $N=4$ Toda (KdV)
hierarchy in $N=4$ superspace. Indeed, let us remember that there exist
another Lax pair formulation of the $N=4$ KdV hierarchy in the $N=2$
superfield basis (a) $\{{\widetilde V}(z,\theta^+,\eta^+),
F(z,\theta^+,\eta^+), {\overline F}(z,\theta^+,\eta^+)\}$ with the second
flow equations \p{releqa} \cite{dg,ik}. The corresponding Lax
operator ${\cal L}$ is
\begin{eqnarray}
{\cal L} = {\overline {\cal D}}^{+}{\cal D}_+{\partial}^{-1} ~
\Big({\partial}-2{\widetilde V}-
F{\partial}^{-1}{\overline F}\Big) ~{\overline {\cal D}}^{+}{\cal D}_+
{\partial}^{-1}.
\label{lax5}
\end{eqnarray}
We introduce a new
superfield basis $\{V(z,\theta^+,\eta^+), {\overline V}(z,\theta^+,\eta^+)\}$
inspired by the formulae (\ref{c}--\ref{invc}) of the basis (c),
\begin{eqnarray}
V\equiv \frac{1}{2}(F+{\overline F})-i{\widetilde V}, \quad
{\overline V}\equiv \frac{1}{2}(F+{\overline F})+i{\widetilde V},
\label{c1}
\end{eqnarray}
\begin{eqnarray}
{\widetilde V} \equiv \frac{i}{2}(V-{\overline V}), \quad
F \equiv {\cal D}_+{\overline {\cal D}}^{+}{\partial}^{-1}
({\overline V}+V), \quad  {\overline F} \equiv
{\overline {\cal D}}^{+}{\cal D}_+{\partial}^{-1}({\overline V}+V),
\label{invc1}
\end{eqnarray}
\begin{eqnarray}
{\cal L}= {\overline {\cal D}}^{+}{\cal D}_+{\partial}^{-1}
~\Big({\partial}-i(V-{\overline V})-
[{\cal D}_+{\overline {\cal D}}^{+}{\partial}^{-1}({\overline V}+V)]
{\partial}^{-1}[{\overline {\cal D}}^{+}{\cal D}_+{\partial}^{-1}
({\overline V}+V)]\Big)~{\overline {\cal D}}^{+}{\cal D}_+{\partial}^{-1}.
\label{lax6}
\end{eqnarray}
Finally, in analogy with our derivation of the Lax operator $L_1$
in \p{lax3}, we replace the $N=2$ superfields $V$ and ${\overline V}$
in \p{lax5} by one chiral ${\cal V}(z,\theta^+,\theta^-,\eta^+,\eta^-)$
and one antichiral
${\overline {\cal V}}(z,\theta^+,\theta^-,\eta^+,\eta^-)$
even $N=4$ superfield,
\begin{eqnarray}
\hbox{\bbd D}_{\pm}{\cal V} =0, \quad
{\overline {\hbox{\bbd D}}}^{\pm}~{\overline {\cal V}} = 0.
\label{N=4constrnew}
\end{eqnarray}
One can check that the resulting Lax pair representation
\begin{eqnarray}
{(-1)}^{l+1}{\textstyle{\partial\over\partial t_l}}{\cal L}_1 =
[({{\cal L}_1}^{l})_{+} ,{\cal L}_1]
\label{laxreprnew}
\end{eqnarray}
with the new $N=4$ Lax operator ${\cal L}_1$,
\begin{eqnarray}
{\cal L}_1 = {\overline {\cal D}}^{+}{\cal D}_+{\partial}^{-1}
~\Big({\partial}-i({\cal V}-{\overline {\cal V}})-
[{\cal D}_+{\overline {\cal D}}^{+}{\partial}^{-1}
({\overline {\cal V}}+{\cal V})]
{\partial}^{-1}[{\overline {\cal D}}^{+}{\cal D}_+{\partial}^{-1}
({\overline {\cal V}}+{\cal V})]\Big)~{\overline {\cal D}}^{+}
{\cal D}_+{\partial}^{-1},
\label{lax7}
\end{eqnarray}
gives consistent $N=4$ supersymmetric flows
${\textstyle{\partial\over\partial t_l}}$ as well. Here, we have
introduced a new superspace basis
$\{ {\cal D}_{\pm}, {{\overline {\cal D}}}^{\pm} \}$  $\Rightarrow$
$\{ {\hbox{\bbd D}}_{\pm}, {\overline {\hbox{\bbd D}}}^{\pm}\}$
defined by
\begin{eqnarray}
&& \quad  \quad  \quad {\cal D}_+\equiv
\frac{1}{\sqrt{2}}({\overline {\hbox{\bbd D}}}^{-}+{\hbox{\bbd D}}_{+}),
\quad {\overline {\cal D}}^{-}\equiv
\frac{1}{\sqrt{2}}({\overline {\hbox{\bbd D}}}^{-}-{\hbox{\bbd D}}_{+}),
\nonumber\\
&&\quad  \quad  \quad {\cal D}_-\equiv
\frac{1}{\sqrt{2}}(
{\hbox{\bbd D}}_{-}-{\overline {\hbox{\bbd D}}}^{+}),\quad
{\overline {\cal D}}^{+}\equiv
\frac{1}{\sqrt{2}}({\hbox{\bbd D}}_{-}+
{\overline {\hbox{\bbd D}}}^{+}),\nonumber\\
&&\Bigl\{{\hbox{\bbd D}}_{k}\,,\,
{\overline {\hbox{\bbd D}}}^{m}\Bigr\}=
{{\delta}_{k}}^{m} {\partial}, \quad
\Bigl\{{\hbox{\bbd D}}_{k}\,,\,{\hbox{\bbd D}}_{m}\Bigr\}=
\Bigl\{{\overline {\hbox{\bbd D}}}^{k}\,,\,
{\overline {\hbox{\bbd D}}}^{m}\Bigr\}=0,\quad k,m=\pm.
\label{new1}
\end{eqnarray}

As an example, let us present explicitly the second flow equations
\begin{eqnarray}
&&i{\textstyle{\partial\over\partial t_2}}{\cal V} =
(+i{\hbox{\bbd D}}_{-}{\hbox{\bbd D}}_{+}{\overline {\cal V}}
+\frac{1}{2}({\hbox{\bbd D}}_{-}{\hbox{\bbd D}}_{+}
{\partial}^{-1}{\overline {\cal V}})^2-\frac{3}{2} {\cal V}^2)~'-
{\hbox{\bbd D}}_{-}{\hbox{\bbd D}}_{+}
({\overline {\cal V}}~{\overline {\hbox{\bbd D}}}^{-}
{\overline {\hbox{\bbd D}}}^{+}{\partial}^{-1}{\cal V}), \nonumber\\
&&i{\textstyle{\partial\over\partial t_2}}{\overline {\cal V}} =
(-i{\overline {\hbox{\bbd D}}}^{-}{\overline {\hbox{\bbd D}}}^{+}{\cal V}
-\frac{1}{2}({\overline {\hbox{\bbd D}}}^{-}
{\overline {\hbox{\bbd D}}}^{+}
{\partial}^{-1}{\cal V})^2+\frac{3}{2}{\overline {\cal V}}^2)~'+
{\overline {\hbox{\bbd D}}}^{-}{\overline {\hbox{\bbd D}}}^{+}
({\cal V}~{\hbox{\bbd D}}_{-}{\hbox{\bbd D}}_{+}
{\partial}^{-1} {\overline {\cal V}})
\label{eqs2jN=4new}
\end{eqnarray}
resulting from equations (\ref{laxreprnew}--\ref{lax7}).

Let us analyze these equations in the same way as it has been
done in the previous section. We introduce the
$N=2$ superfield basis $\{{\widetilde {\cal V}}(z,\theta^+,\eta^-),
{\cal \phi}(z,\theta^+,\eta^-),
{\overline {\cal \phi}}(z,\theta^+,\eta^-)\}$
\begin{eqnarray}
&& {\widetilde {\cal V}}\equiv
\frac{i}{2}(({\hbox{\bbd D}}_{-}+{\overline {\hbox{\bbd D}}}^{-})
({\hbox{\bbd D}}_{+}+{\overline {\hbox{\bbd D}}}^{+})^{-1}
({\overline {\cal V}} - {\cal V}))
(z,\theta^+,\theta^-=-{\theta}^{+},\eta^+, \eta^-=\eta^{+}), \nonumber\\
&&{\cal \phi}\equiv {\cal V}
(z,\theta^+,\theta^-=-{\theta}^{+},\eta^+, \eta^-=\eta^{+}), \quad
\hbox{\bbd D}_{+}~{\cal \phi} =0, \nonumber\\
&&{\overline {\cal \phi}}\equiv
{\overline {\cal V}}
(z,\theta^+,\theta^-=-{\theta}^{+},\eta^+, \eta^-=\eta^{+}), \quad
{\overline {\hbox{\bbd D}}}^{+} ~{\overline {\cal \phi}} = 0
\label{anew}
\end{eqnarray}
which is similar to the basis (a) \p{a}.
Then one observes that the corresponding second flow equations
reproduce the second flow equations \p{releqc} of the basis (c) with the
following correspondence:
\begin{eqnarray}
\{{\widetilde J},\Phi, {\overline \Phi},D, {\overline D},
{\textstyle{\partial\over\partial t_2}}\}
\Rightarrow
\{{\widetilde {\cal V}},{\cal \phi}, {\overline {\cal \phi}},
{\hbox{\bbd D}}_{+},{\overline {\hbox{\bbd D}}}^{+},
-{\textstyle{\partial\over\partial t_2}}\}.
\label{bnew}
\end{eqnarray}
Therefore, we conclude that the $N=4$ superfield
equations \p{eqs2jN=4} and \p{eqs2jN=4new} are the second flow
equations of the same $N=4$ supersymmetric Toda (KdV) hierarchy
written in two different $N=4$ superfield bases,
$\{{\cal J},{\overline {\cal J}}, {\cal D}_{\pm},
{\overline {\cal D}}_{\pm}\}$ and
$\{{\cal V},{\overline {\cal V}}, {\hbox{\bbd D}}^{\pm},
{\overline {\hbox{\bbd D}}}^{\pm}\}$, respectively.
The explicit relation between these two bases is the following:
\begin{eqnarray}
&&2i{\cal V}={\overline {\cal J}}-{\cal J}+
({\cal D}_{-}+{\overline {\cal D}}^-)
({\cal D}_{+}+{\overline {\cal D}}^+)^{-1}
({\overline {\cal J}}+{\cal J}), \nonumber\\
&&2i{\overline {\cal V}}={\overline {\cal J}}-{\cal J}-
({\cal D}_{-}+{\overline {\cal D}}^-)
({\cal D}_{+}+{\overline {\cal D}}^+)^{-1}
({\overline {\cal J}}+{\cal J}), \nonumber\\
&&{\hbox{\bbd D}}_{+}=\frac{1}{\sqrt{2}}({\cal D}_+ -
{\overline {\cal D}}^-), \quad
{\overline {\hbox{\bbd D}}}^{-}=\frac{1}{\sqrt{2}}({\cal D}_+ +
{\overline {\cal D}}^-), \nonumber\\
&&{\hbox{\bbd D}}_{-}=\frac{1}{\sqrt{2}}({\overline {\cal D}}^+ +
{\cal D}_-), \quad
{\overline {\hbox{\bbd D}}}^{+}=\frac{1}{\sqrt{2}}({\overline {\cal D}}^+
- {\cal D}_-).
\label{relnewlast}
\end{eqnarray}

Actually, the Lax pair representation  (\ref{laxreprnew}--\ref{lax7}) is
a direct consequence of the Lax pair representation
(\ref{laxreprnew2}--\ref{newlax}). In order to show this, one
can first express the Lax operator ${\cal L}_1$ \p{lax7} in terms of the
superfields ${\cal J}$ and  ${\overline {\cal J}}$ using equations
\p{relnewlast}, then verify that it is
the square of the Lax operator ${\hbox{\bbd L}}$ \p{newlax},
\begin{eqnarray}
{\cal L}_1 &=& {\overline {\cal D}}^{+}{\cal D}_+{\partial}^{-1}
~\Big({\partial} - [({\cal D}_- + {\overline {\cal D}}^{-})
({\cal D}_+ + {\overline {\cal D}}^{+})^{-1}
({\overline {\cal J}}+{\cal J})]-
{\cal J}{\partial}^{-1}{\overline {\cal J}}\Big)
~{\overline {\cal D}}^{+}{\cal D}_+{\partial}^{-1}\nonumber\\
& \equiv & \Big(
~{\overline {\cal D}}^{+}{\cal D}_+{\partial}^{-1}
~\Big(-{\cal D}_- - {\overline {\cal D}}^{-}+
[({\cal D}_+ + {\overline {\cal D}}^{+})^{-1}
({\overline {\cal J}}+{\cal J})]\Big)
~{\overline {\cal D}}^{+}{\cal D}_+{\partial}^{-1}~\Big)^2
\equiv {\hbox{\bbd L}}^2.
\label{laxnew7}
\end{eqnarray}

If one introduces the notation
\begin{eqnarray}
&&{\hbox{\bbd V}}^{12} \equiv {\hbox{\bbd V}}^{21} \equiv
\frac{1}{2}({\hbox{\bbd D}}_{-}+{\overline {\hbox{\bbd D}}}^{-})
({\hbox{\bbd D}}_{+}+{\overline {\hbox{\bbd D}}}^{+})^{-1}
({\overline {\cal V}} - {\cal V}),
\label{defnew}
\end{eqnarray}
\begin{eqnarray}
{\hbox{\bbd V}}^{11}\equiv {\overline {\cal V}}, \quad
{\hbox{\bbd V}}^{22}\equiv {\cal V}, \quad
{\hbox{\bbd D}}_1 \equiv {\hbox{\bbd D}}_{+}, \quad
{\hbox{\bbd D}}_2 \equiv  {\overline {\hbox{\bbd D}}}^{-}, \quad
{\overline {\hbox{\bbd D}}}^1 \equiv  {\overline {\hbox{\bbd D}}}^{+}, \quad
{\overline {\hbox{\bbd D}}}^2 \equiv  {\hbox{\bbd D}}_{-},
\label{defnew1}
\end{eqnarray}
then the chirality constraints \p{N=4constrnew} and
relations \p{defnew} can equivalently be rewritten as
\begin{eqnarray}
{\hbox{\bbd D}}^{(i}{\hbox{\bbd V}}^{jk)} = 0, \quad
{\overline {\hbox{\bbd D}}}^{(i}{\hbox{\bbd V}}^{jk)}= 0,
\quad i,j,k = 1,2
\label{defnew2}
\end{eqnarray}
which is similar to the formulae (\ref{def}--\ref{def2}). In terms
of the objects $\{V^{ij}, {\cal D}^{i},{\overline {\cal D}}^{i}\}$ and
$\{{\hbox{\bbd V}}^{ij},{\hbox{\bbd D}}^{i},
{\overline {\hbox{\bbd D}}}^{i}\}$ the transformation
\p{relnewlast} is the $SU(2)$ rotation
\begin{eqnarray}
{\hbox{\bbd D}}^{i}= (S {\cal D})^{i}, \quad
{\overline {\hbox{\bbd D}}}^{i}= (S {\overline {\cal D}})^{i}, \quad
{\hbox{\bbd V}}^{ij} = {(SVS^{T})}^{ij}, \quad
S=\frac{1}{\sqrt{2}}\left(\begin{array}{cc} +1 & -1 \\ +1 & +1
\end{array}\right),
\label{defnew3}
\end{eqnarray}
where the matrix $S \in SU(2)$ and $S^{T}$ is the transposed matrix.

\section{Conclusion}

In this paper, we have developed a consistent Lax pair formulation
of the $N=4$ supersymmetric Toda chain (KdV) hierarchy in $N=4$
superspace. The explicit general
formulae \p{flowsform} for its bosonic flows in terms of the Lax operator
in $N=4$ superspace have been derived. Then, a change of basis in $N=4$
superspace has allowed us to eliminate all nonlocalities in the flows
(\ref{basisnew4}--\ref{b4}). We have also explicitly presented
the general formula for the corresponding Hamiltonians \p{hamiltonians},
the first two Hamiltonian structures (\ref{hameq1}--\ref{hameq2}) as well as
the recursion operator (\ref{recop0}-\ref{recrel}) in $N=4$ superspace.
Then, we have simplified these formulae by rewriting them in terms of the
single real linear $N=4$ superfield ${\xi}$
(\ref{hamstrxi1}--\ref{recopxi}) whose flows are local. Furthermore,
we have established its relationship \p{relations} to the $N=4$ $O(4)$
superconformal  supercurrent which hopefully could shed light on the
longstanding problem of constructing the $N=4$ $O(4)$ KdV hierarchy. Then,
we have shown that all these flows possess five complex conjugations
(\ref{conj1}--\ref{conj3}) and (\ref{conj1new}--\ref{conj2new})
as well as an infinite-dimensional group of discrete Darboux symmetries
\p{conj2jn=4}. Finally, the explicit formulae
(\ref{def}--\ref{def1}) and (\ref{defnew}--\ref{defnew1})
relating the two different descriptions of the
flows in $N=4$ superspace used in \cite{dik} and \cite{ds} have
established. It is obvious that there remain a
lot of work to do in order to improve our understanding of the hierarchy
in $N=4$ superspace, but the construction of the $N=4$ Lax pair formulation
presented in this paper is a crucial, necessary step towards a
complete description.

{}~

{}~

\noindent{\bf Acknowledgments.} A.S. would like to thank the
Laboratoire de Physique de l'ENS Lyon for the hospitality during
the course of this work. This work was partially supported by the
PICS Project No. 593, RFBR-CNRS Grant No. 01-02-22005, Nato Grant
No. PST.CLG 974874 and RFBR Grant No. 99-02-18417.


\begin{thebibliography}{**}
\bibitem{dgs}
F. Delduc, L. Gallot and A. Sorin,
{\it $N=2$ local and $N=4$ nonlocal reductions of supersymmetric KP
hierarchy in $N=2$ superspace}, Nucl. Phys. {\bf B558} (1999) 545,
solv-int/9907004.
\bibitem{di}
F. Delduc and E. Ivanov, {\it $N=4$ super KdV equation},\\
Phys. Lett. {\bf B309} (1993) 312, hep-th/9301024.
\bibitem{dik}
F. Delduc, E. Ivanov and S. Krivonos,
{\it N=4 super KdV hierarchy in N=4 and N=2 superspaces}, J. Math. Phys.
{\bf 37} (1996) 1356; Erratum-ibid. {\bf 38} (1997) 1224,
hep-th/9510033.
\bibitem{ds}
F. Delduc and A. Sorin,
{\it A note on real forms of the complex $N=4$ supersymmetric Toda chain
hierarchy in real $N=2$ and $N=4$ superspaces},
Nucl. Phys. {\bf B577} (2000) 461, solv-int/9911005.
\bibitem{dg}
F. Delduc and L. Gallot, {\it N=2 KP and KdV hierarchies in extended
superspace},\\ Commun. Math. Phys. {\bf 190} (1997) 395, solv-int/9609008.
\bibitem{ik}
E. Ivanov and S. Krivonos, {\it New integrable extensions of N=2 KdV and
Boussinesq hierarchies}, Phys. Lett {\bf A231} (1997) 75, hep-th/9609191.
\bibitem{ls}
A.N. Leznov and A.S. Sorin,
{\it Two-dimensional superintegrable mappings and integrable
hierarchies in the $(2|2)$ superspace},
{\sl Phys. Lett.} {\bf B389} (1996) 494, hep-th/9608166;\\
{\it Integrable mappings and hierarchies in the $(2|2)$ superspace},\\
{\sl Nucl. Phys.} (Proc. Suppl.) {\bf B56} (1997) 258.
\bibitem{ols}
O. Lechtenfeld and A. Sorin,
{\it Fermionic flows and tau function of the N=$(1|1)$ superconformal
Toda lattice hierarchy}, Nucl. Phys. {\bf B557} (1999) 535,
solv-int/9810009.
\bibitem{ff}
B. Fuchssteiner and A.S. Fokas,
{\it Symplectic structures, their B\"acklund transformations
and hereditary symmetries}, Physica {\bf D4} (1981) 47.
\bibitem{sch}
K. Schoutens, {\it $O(N)$-extended superconformal field theory in
superspace},\\ Nucl. Phys. {\bf B295} (1988) 634.
\bibitem{stp}
A. Sevrin, W. Troost and A. Van Proeyen, {\it Superconformal algebras
in two dimensions with $N=4$}, Phys. Lett. {\bf B208} (1988) 447.
\bibitem{ggrs}
S.J. Gates, Jr., M.T. Grisaru, M. Ro${\check{c}}$ek and W. Siegel,
{\it Superspace or one thousand and one lessons in supersymmetry},
the Benjamin/Cummings Publishing Company, Inc, 1983.
%pgs. 58-59.
\bibitem{w}
P. West, {\it Introduction to supersymmetry and supergravity},
extended second edition, World Scientific, 1990.
% pgs. 393-394.
\bibitem{s}
A. Sorin, {\it The discrete symmetry of the N=2 supersymmetric modified
NLS hierarchy},\\ Phys. Lett. {\bf B395} (1997) 218, hep-th/9611148;
{\it Discrete symmetries of the N=2 supersymmetric Generalized
Nonlinear Schroedinger hierarchies}, Phys. Atom. Nucl. {\bf 61} (1998)
1768, solv-int/9701020.
\bibitem{eh}
J. Evans, T. Hollowood, {\it Supersymmetric Toda field theories},
Nucl. Phys. {\bf B352} (1991) 723.
\bibitem{lds}
V.B. Derjagin, A.N. Leznov and A. Sorin, {\it The solution of the
$N=(0|2)$ superconformal f-Toda lattice}, Nucl. Phys. {\bf B527} (1998)
643, solv-int/9803010.
\bibitem{ks3}
S. Krivonos and A. Sorin, {\it Extended N=2 supersymmetric matrix
(1,s)-KdV hierarchies}, Phys. Lett. {\bf A251} (1999) 109, solv-int/9712002.
\bibitem{dg3}
F. Delduc and L. Gallot, {\it A note on the third family of N=2
supersymmetric KdV hierarchies}, J. of Nonlinear Math. Phys.
{\bf 6} (1999) 1, solv-int/9901011.
\end{thebibliography}
\end{document}